\documentclass[sigconf]{acmart}

\usepackage{float}
\usepackage{xcolor,colortbl} 
\usepackage{amsmath,amsfonts} %
\usepackage{graphicx}%
\usepackage{textcomp}%
\usepackage{xcolor}%

\usepackage{array}%
\usepackage{pifont}%

\usepackage{amsthm}%
\usepackage{bm}

\usepackage[linesnumbered,ruled,vlined]{algorithm2e}%
\usepackage{setspace}%

\usepackage{multirow}%
\usepackage{siunitx}
\usepackage{footnote}%
\usepackage[utf8]{inputenc}
\usepackage[english]{babel}

\usepackage{blindtext}

\usepackage{dblfloatfix}%

\usepackage{xcolor,colortbl} 

\usepackage{xurl}

\usepackage{ctable} 
\graphicspath{ {pdf/} }

\usepackage{pifont}

\usepackage{filecontents}

\usepackage{balance}

\newlength{\bibitemsep}
\newlength{\bibparskip}
\let\oldthebibliography\thebibliography
\renewcommand\thebibliography[1]{%
	\oldthebibliography{#1}%
	\setlength{\parskip}{\bibitemsep}%
	\setlength{\itemsep}{\bibparskip}%
}

\AtBeginDocument{%
  \providecommand\BibTeX{{%
    \normalfont B\kern-0.5em{\scshape i\kern-0.25em b}\kern-0.8em\TeX}}}

\copyrightyear{2022}
\acmYear{2022}
\setcopyright{rightsretained}
\acmConference[CCS '22]{Proceedings of the 2022 ACM SIGSAC Conference on Computer and Communications Security}{November 7--11, 2022}{Los Angeles, CA, USA}
\acmBooktitle{Proceedings of the 2022 ACM SIGSAC Conference on Computer and Communications Security (CCS '22), November 7--11, 2022, Los Angeles, CA, USA}
\acmDOI{10.1145/3548606.3560643}
\acmISBN{978-1-4503-9450-5/22/11}

\usepackage{etoolbox}
\makeatletter 
\patchcmd{\maketitle}{\@copyrightpermission}{
 \begin{minipage}{0.3\columnwidth}
 \href{https://creativecommons.org/licenses/by/4.0/}{\includegraphics[width=0.90\textwidth]{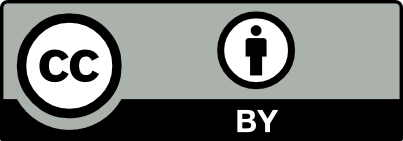}}
 \end{minipage}\hfill
 \begin{minipage}{0.7\columnwidth}
 \href{https://creativecommons.org/licenses/by/4.0/}{This work is licensed under a Creative Commons Attribution International 4.0 License.}
 \end{minipage}

 \vspace{5pt}
}{}{}
\makeatother
	
\begin{document}

\title{
A Wolf in Sheep's Clothing: Spreading Deadly Pathogens Under the Disguise of Popular Music
}

\author{Anomadarshi Barua}
\authornote{Both authors contributed equally to this research.}
\affiliation{%
  \institution{University of California, Irvine}
  \city{Irvine, CA}
  \country{USA}}
\email{anomadab@uci.edu}

\author{Yonatan Gizachew Achamyeleh}
\authornotemark[1]
\affiliation{%
  \institution{University of California, Irvine}
  \city{Irvine, CA}
  \country{USA}}
\email{yachamye@uci.edu}

\author{Mohammad Abdullah Al Faruque}
\affiliation{%
  \institution{University of California, Irvine}
  \city{Irvine, CA}
  \country{USA}}
\email{alfaruqu@uci.edu}

\renewcommand{\shortauthors}{Anomadarshi Barua, Yonatan Gizachew Achamyeleh, \& Mohammad Abdullah Al Faruque}

\begin{abstract}

A Negative Pressure Room (NPR) is an essential requirement by the Bio-Safety Levels (BSLs) in biolabs or infectious-control hospitals to prevent deadly pathogens from being leaked from the facility. An NPR maintains a negative pressure inside with respect to the outside reference space so that microbes are contained inside of an NPR. Nowadays, differential pressure sensors (DPSs) are utilized by the Building Management Systems (BMSs) to control and monitor the negative pressure in an NPR. This paper demonstrates a non-invasive and stealthy attack on NPRs by spoofing a DPS at its resonant frequency. Our contributions are: (1) We show that DPSs used in NPRs typically have resonant frequencies in the audible range. (2) We use this finding to design malicious music to create resonance in DPSs, resulting in an overshooting in the DPS's normal pressure readings. (3) We show how the resonance in DPSs can fool the BMSs so that the NPR turns its negative pressure to a positive one, causing a potential \textit{leak} of deadly microbes from NPRs. We do experiments on 8 DPSs from 5 different manufacturers to evaluate their resonant frequencies considering the sampling tube length and find resonance in 6 DPSs. 
We can achieve a 2.5 Pa change in negative pressure from a $\sim$7 cm distance when a sampling tube is not present and from a $\sim$2.5 cm distance for a 1 m sampling tube length. We also introduce an interval-time variation approach for an adversarial control over the negative pressure and show that the \textit{forged} pressure can be varied within 12 - 33 Pa. Our attack is also capable of attacking multiple NPRs simultaneously. Moreover, we demonstrate our attack at a real-world NPR located in an anonymous bioresearch facility, which is FDA approved and follows CDC guidelines. We also provide countermeasures to prevent the attack.

\vspace{-0.30em}
\end{abstract}

\ccsdesc[500]{Security and privacy~Embedded systems security}
\ccsdesc[300]{Security and privacy~Hardware attacks and countermeasures}


\keywords{Pressure sensors; Resonance; Negative pressure room; Pathogens} 

\maketitle


\thispagestyle{empty}
\vspace{-0.730em}
\section{Introduction}

A Bio-Safety Level (BSL) \cite{ta2019biosafety,risi2010preparing} is a set of strict regulations assigned to a biolab or hospital facility to prevent deadly pathogens from being leaked from the facility. The BSL is ranked from BSL-1 (lowest safety level) to BSL-4 (highest safety level) depending on the microbes that are being contained in a laboratory or hospital setting. The Centers for Disease Control and Prevention (CDC) sets BSLs to exhibit specific controls for the containment of microbes to protect the surrounding environment and community.

BSLs require that the isolation rooms in a biolab or infectious-control hospital maintain negative pressure with respect to the outside hallway \cite{ta2019biosafety}. Therefore, the room is known as the Negative Pressure Room (NPR). An NPR ensures that potentially harmful microbes cannot leak from the facility through airflow by maintaining negative pressure inside. Therefore, an NPR is critical in preventing deadly bioaerosols from escaping from the facility. 

With rising concerns of bioterrorism, an NPR must maintain a certain \textit{negative pressure} following strict regulations established by the CDC, ASHRAE, or other authorities \cite{chinn2003guidelines,paul2008ventilation}. The Differential Pressure Sensors (DPSs) are commonly used in NPRs to measure the negative pressure in the facility \cite{miller2017implementing}. The DPSs provide the pressure data to the Heating, Ventilation, and Air Conditioning (HVAC) systems, which \textit{maintains} the negative pressure by controlling the airflow into NPRs \cite{underwood2002hvac}. In addition, a Room Pressure Monitoring (RPM) system is also present in NPRs to \textit{monitor} the room pressure \cite{SPRM}. The RPM system also depends on the reading from the DPSs installed in an NPR. Both RPM and HVAC systems are connected with the Building Management Systems (BMSs) for automated control and monitoring of the negative pressure in an NPR.


A DPS has an elastic diaphragm working as a pressure force collector. 
Therefore, a DPS can be modeled as a second-order dynamic system with a resonant frequency \cite{wilkinson2009principles}. We demonstrate by thorough experiments that the resonant frequencies of DPSs used in NPRS are typically in the audible range. In addition, we show that the DPS with a sampling tube can be modeled as a Helmholtz resonator, and the resonant frequency of a DPS with a sampling tube still falls within the audible range. This finding is important because an attacker, who has an intention to change the negative pressure in an NPR, may use an audible sound having a resonant frequency to create resonance in a DPS and generate a \textit{forged }pressure to perturb the normal readings of a DPS located in an NPR.

However, a sound having a single-tone resonant frequency will create a "beep"-ish sound, which makes the attack easily identifiable by the authority. Moreover, the HVAC and RPM systems cannot be fooled by a simple resonance in DPS because these systems have a slower response time compared to a resonance. Therefore, a simple resonance in DPS is not enough to turn NPR's negative pressure into a positive pressure to leak airborne pathogens from an NPR.

To solve the above problems, this paper adopts a smart strategy by \textit{disguising} the resonant frequency band inside popular music. The resonant frequencies are inserted as a \textit{segment} into the music for a certain duration in every specific interval. Every inserted segment of the resonant frequency is ended at its peak. Therefore, the corresponding pressure wave inside a DPS also ends at its peak. As a DPS with a sampling tube is a second-order oscillating system \cite{bajsic2007response}, the pressure wave does not instantly fall to zero from the peak value. Instead, the pressure wave starts to attenuate from its peak exponentially. If the interval between two consecutive segments is small, the pressure wave never falls below a certain value. Therefore, a forged pressure is always present inside a DPS having an average value greater than zero. As a result, the malicious music injected into the DPS can fool the controller of HVAC and RPM systems connected with BMSs to change the negative pressure of an NPR into a positive one. Moreover, the segments of resonant frequency are camouflaged in the malicious music so that the attack is not identifiable by the authority. Therefore, we name this attack as "\textit{the wolf in sheep's clothing}" since this strategy ensures stealthiness. 

The consequences of changing a negative pressure into a positive one can be catastrophic. If the NPR has an infectious patient admitted or an ongoing bioresearch, the attacker can control the timing of the attack to \textit{leak} a deadly pathogen \textit{from} the NPR. Moreover, an abnormal change in NPR's pressure triggers an alarm that may create chaos in the facility. An attacker can use this chaos to initiate a stronger attack, such as stealing deadly microbes from the NPR or physically attacking the biosafety cabinets in an NPR. Therefore, our attack model is strong and impactful and has the potential to cause tremendous losses in human lives and monetary resources.

\textbf{Contributions:} We have the following technical contributions:

\textbf{(1)} We evaluate eight industry-used pressure sensors from five different manufacturers to show that the pressure sensors used in NPRs have resonant frequencies in the audible range. 

\vspace{0.1em}
\textbf{(2)} We design malicious music disguising the resonant frequencies of DPSs inside of the music to fool the HVAC and RPM systems of an NPR. We show through experiments that this strategy can change the negative pressure of an NPR to a positive one.

\vspace{0.1em}
\textbf{(3)} We show that the attacker can adversarially control the forged pressure in DPSs by using the malicious music. Moreover, we show that the attacker can also \textit{simultaneously} attack \textit{multiple} NPRs in a facility using our attack model.

\vspace{0.1em}
\textbf{(4)} We demonstrate our attack model at a real-world NPR located in an anonymous bioresearch facility. The NPR is approved by the Food and Drug Administration (FDA) and follows CDC guidelines. We also provide countermeasures to prevent the attack on NPRs.

\textbf{Demonstration:} The demonstration of the attack is shown in the following link: {\color{blue}\url{https://sites.google.com/view/awolfinsheepsclothing/home}}

\vspace{-0.3em}
\section{Background}

\subsection{NPR and its importance}
\label{subsec:NPR and its Importance}

An NPR \cite{tsai2006airborne} maintains lower pressure inside with respect to the outside reference space. As air typically travels from higher pressure areas to lower pressure areas, NPR ensures that clean air is drawn into the room so that contaminated particles inside the room are not able to escape. This is why NPRs are present in hospitals and biosafety labs as they prevent airborne particles like bacteria and viruses from spreading out from the facility. NPRs are also present in safety-critical facilities, such as pharmacies and clean rooms.


\textbf{Importance}: 
The safety of NPRs is paramount as spreading airborne microbes from NPRs may result in catastrophic consequences. For example, a deadly fungus belonging to the genus \textit{Aspergillus} is an airborne pathogen that can cause Aspergillosis disease resulting in acute pneumonia and abscesses of the lungs and kidneys \cite{CDCairborne}. It has a mortality rate of $\sim$100\% for people with neutropenia (i.e., low neutrophils).
Respiratory tract infections, such as influenza, swine flu, and COVID-19, are great examples of airborne pathogens that result in a worldwide pandemic. Recently, a conspiracy theory has been rumored about the leakage of the COVID-19 as bioweapons from a biolab \cite{BBCnewspropaganda}. In this context, \textit{imagine} an attacker with the intention of spreading infectious disease as bioweapons may target NPRs, where either infected patients are admitted for isolation or research is carried out on deadly pathogens. Therefore, the security of NPRs is critical and is regulated with strict guidelines.

\subsection{Regulations for NPRs}
\label{subsec:Regulations for NPRs}

With rising concerns about bioterrorism and emerging infectious diseases, there has been a greater emphasis on the proper regulations of NPRs. NPRs must follow requirements established by the CDC \cite{chinn2003guidelines}, ASHRAE \cite{paul2008ventilation}, and healthcare design construction guidelines \cite{bartley2010current} to correctly manage airborne infections. Different authorities follow their own regulations \cite{taiwanguideline,jensen2005guidelines,USAAIA,australiaguideline} to maintain a certain negative pressure in NPRs (see Table \ref{table:countries_NP_values}). For example, CDC requires that NPRs must maintain a negative pressure differential of at least $\sim$2.5 Pa (i.e., 0.01 inch water column) in a hospital or biolabs and change the air at least $12$ times per hour \cite{chinn2003guidelines}. Moreover, exhaust from NPRs must be allowed to exit directly outside without contaminating exhaust from other locations. In addition, all exhaust air must be discharged through a High-Efficiency Particulate Air (HEPA) filter to prevent any contamination in the environment.

	\begin{table}[h!]
		\footnotesize
		\centering
			\vspace{-0.90em}
			\caption{Regulations for a Negative Pressure Room (NPR).}
			\vspace{-01.60em}
			\label{table:countries_NP_values}
			\begin{tabular}{p{2.65cm}|p{0.7cm}|p{1.15cm}|p{1.05cm}|p{0.9cm}}
				\hline
				 \cellcolor [gray]{0.85} \textbf{Country} & \cellcolor [gray]{0.85} \textbf{Taiwan} & \cellcolor [gray]{0.85} \textbf{CDC(USA)} & \cellcolor [gray]{0.85} \textbf{AIA(USA)} & \cellcolor [gray]{0.85} \textbf{Australia}\\
				\hline
				\hline
				Negative pressure  & -8 Pa & -2.5 Pa & -2.5 Pa & -15 Pa\\
				\hline
				Air change per hour (ACH) & 8 -12 & > 12 & > 12 & > 12\\
				\hline
			\end{tabular}
		\vspace{-1.5em}
	\end{table}

\subsection{Types of pressure sensors used in NPRs}
\label{subsec:Transducer based pressure sensors (TBPSs)}

Traditionally, hot-wire anemometers \cite{corrsin1947extended} and ball pressure sensors \cite{ballpressureindicator} were used to measure pressure in NPRs. However, they have limitations, such as they are highly sensitive to dust, require periodic maintenance, and cannot be connected to a BMS or RPM for real-time control. Therefore, transducer-based pressure sensors (TBPSs) are replacing hot-wire and ball pressure sensors in NPRs since TBPSs are more accurate, reliable, require low maintenance, and can be connected to BMS or RPM for real-time monitoring. 

\textbf{Physics of TBPSs:} A force collector and a transducer are two fundamental components of TBPSs. A force collector, such as an elastic diaphragm, is combined with a transducer to generate an electrical signal \cite{abacus2021pressure} proportional to the input pressure.

\begin{figure}[h!]
\vspace{-0.60em}
\centering
\includegraphics[width=0.4\textwidth,height=0.11\textheight]{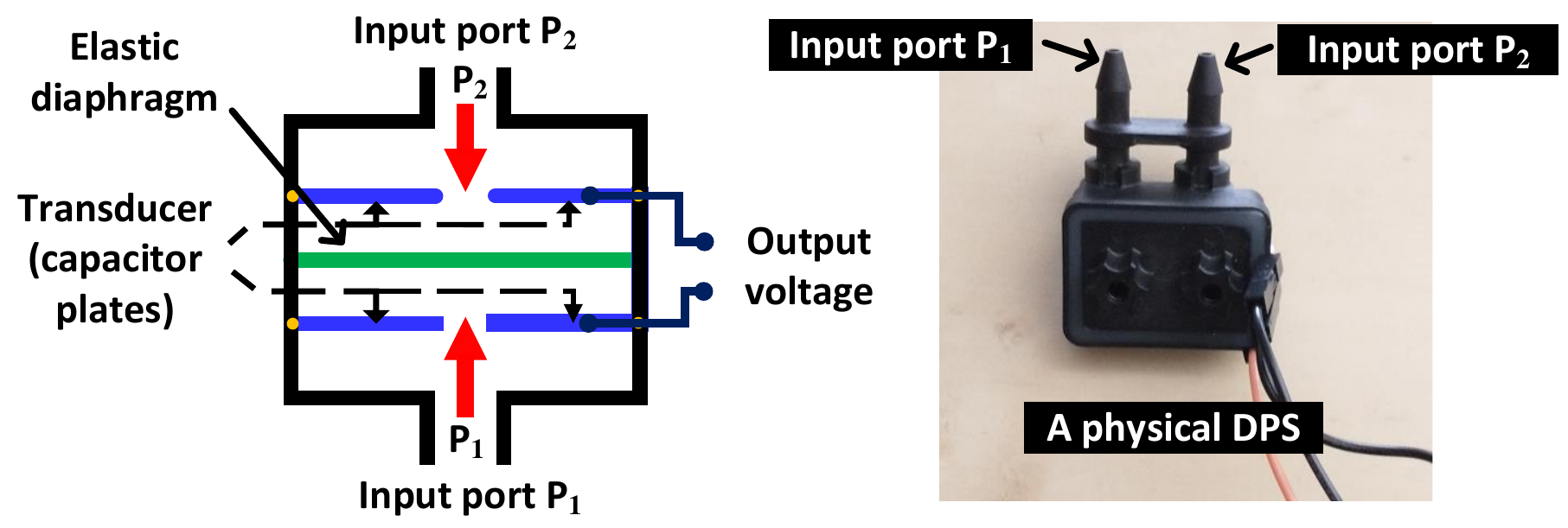}
\vspace{-1.10em}
\caption{Basics of a DPS having two input ports.}
\label{fig:DPS_NPR}
\vspace{-01.70em}
\end{figure}

\textbf{Types of TBPSs:} In general, TBPSs work in one of three modes: absolute, gauge, or differential measurement. 
Absolute pressure sensors use vacuum pressure, and gauge sensors use local atmospheric pressure as the static reference pressure. On the other hand, \textbf{Differential Pressure Sensors (DPSs)} measure the difference between any two pressure levels using two input ports (see Fig. \ref{fig:DPS_NPR}). Therefore, DPSs are naturally suitable in such applications where the \textit{pressure difference} is required to be measured, such as in NPRs \cite{Pressure_Sensing}. 
As a DPS has \textit{high sensitivity} to differential pressure and is deployed in NPRs, we focus on DPSs in next sections.

\subsection{Types of differential pressure sensors}
\label{subsec:Types of DPSs based on transducers}

DPSs typically have a elastic diaphragm placed in between two pressure input ports $P_1$ and $P_2$ (see Fig. \ref{fig:DPS_NPR}). The diaphragm senses the differential pressure $P_1$ - $P_2$ applied to the pressure input ports by changing its shape. The diaphragm's shape change is converted to a proportional output voltage by using a transducer. DPSs either use a \textit{capacitor}, or a \textit{piezoresistor}, or \textit{thermal mass-flow} as a transducer. A DPS is named after the type of transducer it has. 

Fig. \ref{fig:DPS_NPR} shows a capacitive DPS as an example. The diaphragm is placed in between rigid capacitor plates. A differential pressure applied to the diaphragm generates a proportional change in the capacitive transducer resulting in a proportional voltage at the sensor output. We refer to Appendix \ref{appendix:Types of DPSs} for details on other types.

\subsection{Differential pressure sensors used in NPRs}
\label{subsec:DPSs used in NPRs}


DPSs are highly sensitive to a small differential change in the low pressure range (i.e., Pa range) and are naturally suitable to measure a pressure difference. Therefore, DPSs are a \textit{natural choice} to be used in most RPM/BMS systems to control the negative pressure. To prove the prevalence of DPSs in NPRs, we investigate six industry-used RPM systems designed by popular manufacturers. All of these RPM systems use different types of DPSs that are shown in Table \ref{table:DPS in NPR}. 

\begin{table}[h!]
\vspace{-0.920em}
	\footnotesize
	\centering
		\caption{Differential pressure sensors used in NPRs}
		\vspace{-01.30em}
		\label{table:DPS in NPR}
		\begin{tabular}{p{0.2cm}|p{2.3cm}|p{1.35cm}|p{1.20cm}|p{1.40cm}}
			\hline
			 \cellcolor [gray]{0.85} \textbf{Sl.} & \cellcolor [gray]{0.85} \textbf{RPM/DPS part\#} & \cellcolor [gray]{0.85} \textbf{Type} &  \cellcolor [gray]{0.85} \textbf{Technology} &  \cellcolor [gray]{0.85} \textbf{Manufacturer} \\
			\hline
			\hline
			1 & Series RSME \cite{DwyerRSM-1-A} & Capacitive & Differential & Dwyer \\
			\hline
    		2 & SRPM 0R1WB \cite{SPRM} & Capacitive & Differential & Setra  \\
			\hline
		    3 & One Vue Sense \cite{primex} & Unknown  & Differential & Primex  \\
			\hline
			4 & RSME-B-003 \cite{RSMEB003} & Piezoresistive   & Differential & Dwyer  \\
			\hline
			5 & Siemens 547-101A \cite{Siemens} & Unknown & Differential & Siemens \\
			\hline
			6 & Series A1 \cite{Sensocon} & Piezoresistive & Differential & Sensocon \\
			\hline
			7 & GUARDIAN \cite{guardian} & Unknown & Differential & Paragon Con. \\
			\hline
		\end{tabular}
	\vspace{-1.600em}
\end{table}
\vspace{-0em}

\subsection{Resonant frequency of a DPS and resonance} 
\label{subsec:Resonant frequency of a DPS and resonance}

\,\,\,\,\,\,\textbf{Resonant frequency:} As mentioned in Section \ref{subsec:Transducer based pressure sensors (TBPSs)} and \ref{subsec:Types of DPSs based on transducers}, typically, DPSs have a diaphragm/membrane and a transducer. Therefore, the pressure transducer system in DPS is considered as a second-order dynamic system, analogous to a bouncing ball \cite{wilkinson2009principles}. Hence, the transducer system in a DPS has its own resonant frequency, $f_r$, which depends on the mass and stiffness of the diaphragm and mass of the pressure medium as Eqn. \ref{eqn:resonant_frequency} \cite{secondordersystem}.

\vspace{-001.000em}
\begin{equation}
\begin{aligned}
 f_r = \frac{1}{2 \pi} \sqrt{\frac{\text{stiffness of a diaphragm}}{\text{mass of the pressure medium and diaphragm}}} 
\vspace{-00.960em}
\label{eqn:resonant_frequency}
\end{aligned}
\end{equation}

\begin{figure}[h!]
\vspace{-1.10em}
\centering
\includegraphics[width=0.26\textwidth,height=0.1\textheight]{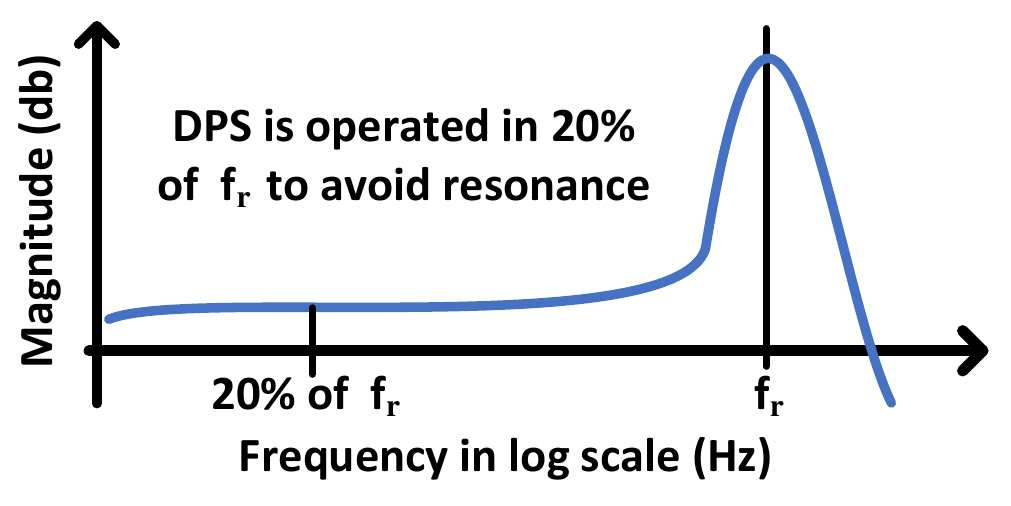}
\vspace{-1.5em}
\caption{Resonant frequency in a DPS.}
\label{fig:resonant_frequency}
\vspace{-01.70em}
\end{figure}

\textbf{Resonance:} Resonance occurs when the frequency of the input pressure wave matches the resonant frequency of the driven transducer system in a DPS, resulting in oscillations \cite{halliday2013fundamentals} in the transducer at large amplitude. This results in significant error by overshooting the peaks and troughs in the actual pressure wave, with an overestimation/underestimation of the actual reading. Therefore, users ensure that a DPS typically operates below its resonant frequency to prevent the resonance. A thumb's rule is 20\% of the resonant frequency is typically used as the usable frequency limit for a given DPS \cite{frequencyresponse}. This concept is illustrated in Fig. \ref{fig:resonant_frequency}.

\subsection{Electronics inside of a DPS}
\label{subsec:Electronics inside of a DPS}

\begin{figure}[h!]
\centering
\includegraphics[width=0.45\textwidth,height=0.11\textheight]{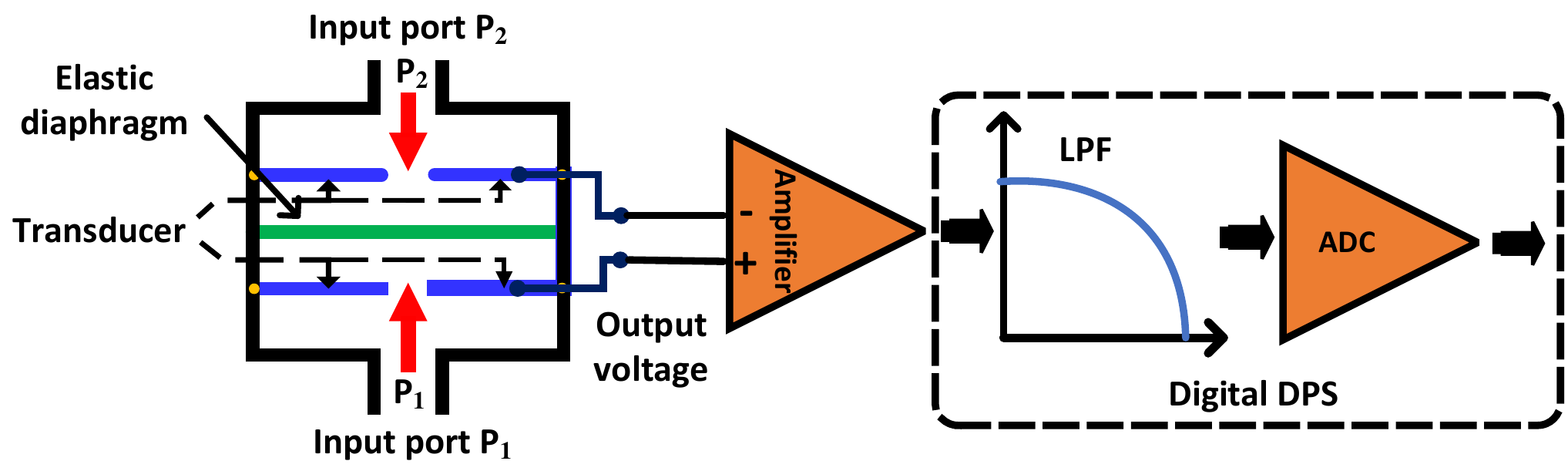}
\vspace{-1.0em}
\caption{Different components inside of a DPS.}
\label{fig:DPS_electronics}
\vspace{-01.40em}
\end{figure}

DPSs have a signal conditioning block in addition to a transducer (see Fig. \ref{fig:DPS_electronics}). The signal conditioning block has differential amplifiers, low-pass filters (LPFs), and analog-to-digital converters (ADCs). A differential amplifier amplifies the output after removing the common-mode noises. An LPF with an ADC digitizes the measured value. Both analog and digital DPSs are available on the market. Analog DPSs output the analog signals from the differential amplifier directly, while digital DPSs contain the LPF and ADC. 

\vspace{-1.0em}

\vspace{-0.30em}
\section{Basics of an NPR}
\label{sec:Basics of an NPR}

This section explains the construction of an NPR, where and how the DPSs are deployed in an NPR, and how the output from the DPS controls the NPR's control system. 

\vspace{-0.40em}
\subsection{Components of a real-world NPR}
\label{subsec:Construction of an NPR}

The components of an NPR vary depending upon the requirements of different facilities. However, the core components are more or less the same for most NPRs. Here, we detail the components of an anonymous NPR where we have visited and experimented with to validate our attack model. \textbf{Please note that the target NPR evaluated in this paper is located in a clean room in an anonymous bioresearch facility. This NPR is also approved by the FDA and follows CDC guidelines.}

\begin{figure}[h]
\vspace{-001.0em}
\centering
\includegraphics[width=0.48\textwidth,height=0.16\textheight]{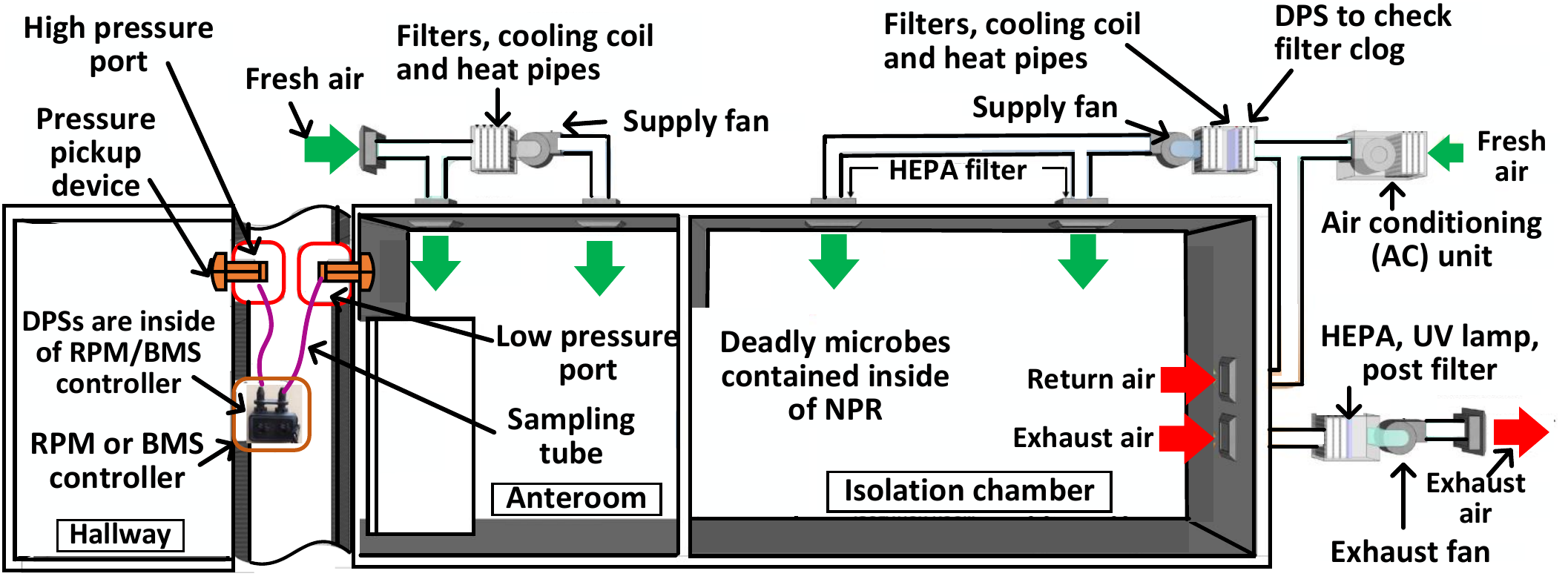}
\vspace{-2.30em}
\caption{Different components of a real-world NPR.}
\label{fig:NPR_construction}
\vspace{-1.100em}
\end{figure}

A typical construction of an NPR is shown in Fig. \ref{fig:NPR_construction}. An NPR has an HVAC system, which includes fresh air inlet ports. The fresh air from the outside is treated with multistage filters and then supplied to the isolation chamber of an NPR, including the anteroom, through an air conditioning (AC) unit. The AC has a Variable Air Volume (VAV) controller, which can increase or decrease
the \textit{supply} fan speed, controlling the fresh airflow to the NPR. 
An exhaust fan continuously moves the contaminated air out from the NPR through a HEPA filter using an exhaust pipe. The polluted air is further treated with a post-filtration unit having an Ultraviolet (UV) lamp. The room is maintained as airtight as possible. An RPM system is installed at the wall and integrated with the BMSs. 

\vspace{-0.50em}
\subsection{How DPSs are deployed in an NPR}
\label{subsec:How DPSs are deployed in an NPR}

\begin{figure}[h!]
\vspace{-0.750em}
\centering
\includegraphics[width=0.48\textwidth,height=0.14\textheight]{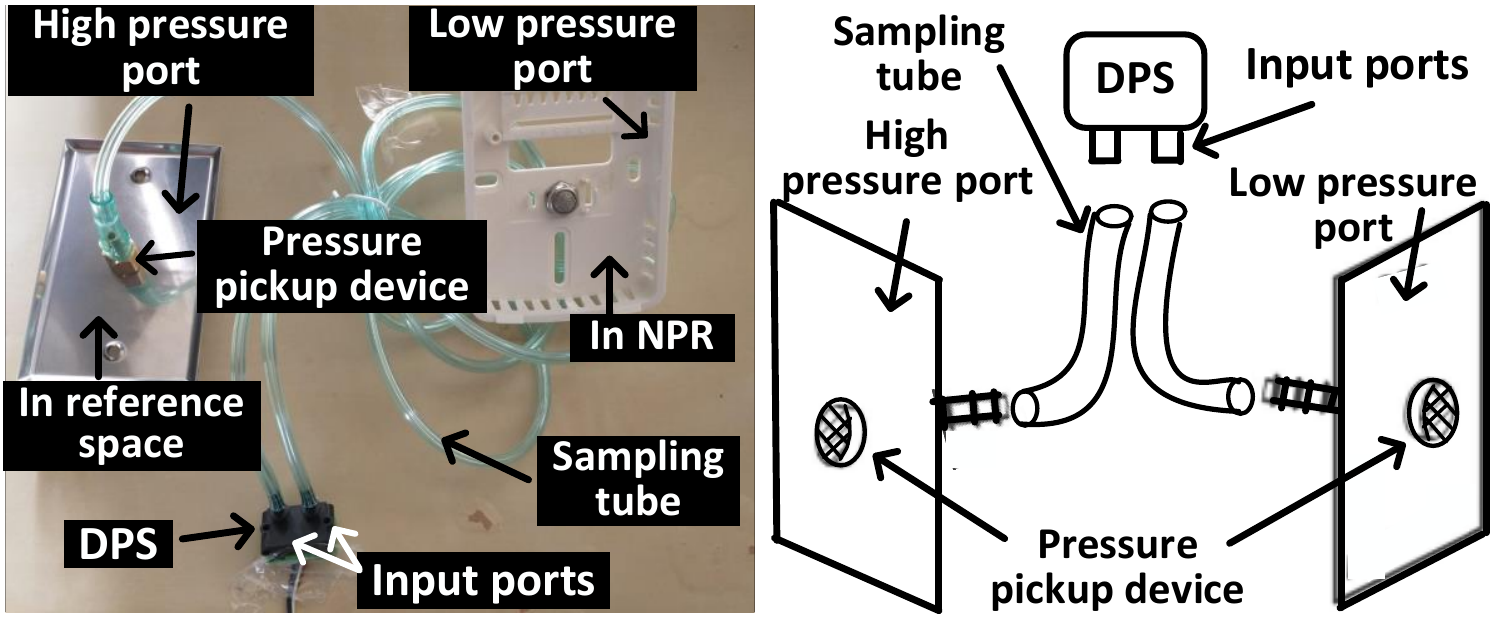}
\vspace{-2.2em}
\caption{Pressure ports and sampling tube of a DPS.}
\label{fig:DPS_pressure ports}
\vspace{-0.0em}
\end{figure}

The HVAC system \textit{ensures} a negative pressure in the NPR by controlling the fresh air and exhaust airflow using the supply and exhaust fan. An RPM system continuously \textit{monitors} the negative room pressure. The RPM and HVAC systems use DPSs to monitor and control negative pressure in an NPR. 
The DPS is typically located inside of RPM or BMS controller. Commonly, the input ports of a DPS are connected with pressure ports using sampling tubes (see Fig. \ref{fig:NPR_construction} and \ref{fig:DPS_pressure ports}). The pressure port located inside an NPR is known as a \textit{low pressure port}. The pressure port located outside an NPR in a hallway/reference space is known as a \textit{high pressure port}. The sampling tube is connected with a pressure pickup device in the pressure ports. The pressure pick-up device increases the surface area of the sampling tube to pick up the target pressure accurately. 

The low and high pressure ports are \textit{exposed} and typically installed in \textit{eyesight} near the door wall or on the ceiling of an NPR. There are other DPSs used in the HVAC system to indicate whether the filters of the HVAC are clogged or not. Typically they are not installed in the eyesight. Therefore, they are not accessible. 

\vspace{-0.2em}
\subsection{Pressure control algorithm in an NPR}
\label{subsec:Control algorithm of an NPR}

A pressure control algorithm running on the BMS controls the HVAC system of the NPR to maintain a constant negative pressure. A simplified control algorithm \ref{alg:controlAlgorithm} is provided below. Algorithm \ref{alg:controlAlgorithm} shows that 
the pressure readings from DPSs are used to control the speed of the supply fan and exhaust fan when the negative pressure increases or decreases from a reference value in the NPR, maintaining the negative pressure close to the reference value. The rest of the control algorithm \ref{alg:controlAlgorithm} is self-explanatory.


\vspace{-00.800em}
\setlength{\textfloatsep}{0pt}
\begin{algorithm}[ht!] 
    \setstretch{1}
	\footnotesize
	\DontPrintSemicolon
	\caption{Pressure control algorithm in an NPR.}
	\label{alg:controlAlgorithm}
	\KwIn{Pressure measurement data from DPSs
	     }
	\KwOut{Send control signals to the HVAC system 
	      } 
	\For{$t \gets 1$ \textbf{to} $\infty$} { 
		Track differential pressure reading from DPS's pressure ports\\
		\If{Negative differential pressure increases from a reference value}{
            Reduce the supply fan speed of the AC to control the fresh airflow\\
            Increase the exhaust fan speed to increase the exhaust airflow\\
		}
		\ElseIf{Negative differential pressure decreases from a reference value}{
		    Increase the supply fan speed of the AC to control the fresh airflow\\
            Reduce the exhaust fan speed to reduce the exhaust airflow\\
		}
		\Else{
		    Maintain the same state of the controller
		}
	}
	\vspace{-00.300em}
\end{algorithm}

\vspace{-00.8300em}
\section{Attack Model}

\begin{figure*}[h]
\vspace{-00.70em}
\centering
\includegraphics[width=0.92\textwidth,height=0.15\textheight]{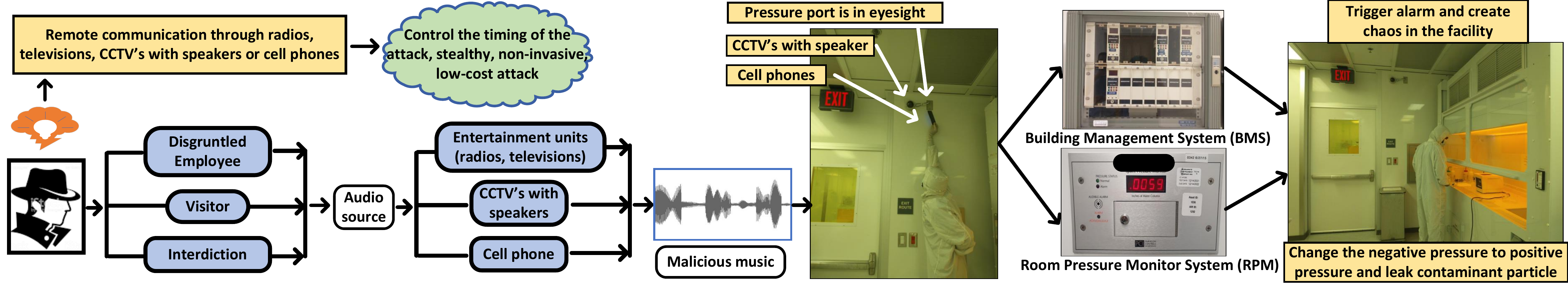}
\vspace{-1.10em}
\caption{A brief overview of the attack model - A Wolf in Sheep's Clothing.}
\label{fig:attack_model}
\vspace{-1.200em}
\end{figure*}

Fig. \ref{fig:attack_model} shows the different components of our attack model associated with NPRs. We discuss the components of the attack model below in a point-by-point fashion.

\textbf{Attacker's intent:} The attacker creates a forged resonance in the DPSs used in NPRs with malicious music having a frequency equal to the resonant frequency of the DPSs. As a result. the overshooting occurs in the actual pressure reading, resulting in a change in the negative pressure maintained in NPRs by the BMSs.

\textbf{Target system:} The attacker targets a facility where NPRs are used to contain deadly microbes and infectious airborne particles. Such facilities include isolation rooms, clean rooms and pharmacies in infectious-control hospitals, and biolabs in bioresearch facilities.

\begin{figure}[h]
\vspace{-00.70em}
\centering
\includegraphics[width=0.45\textwidth,height=0.13\textheight]{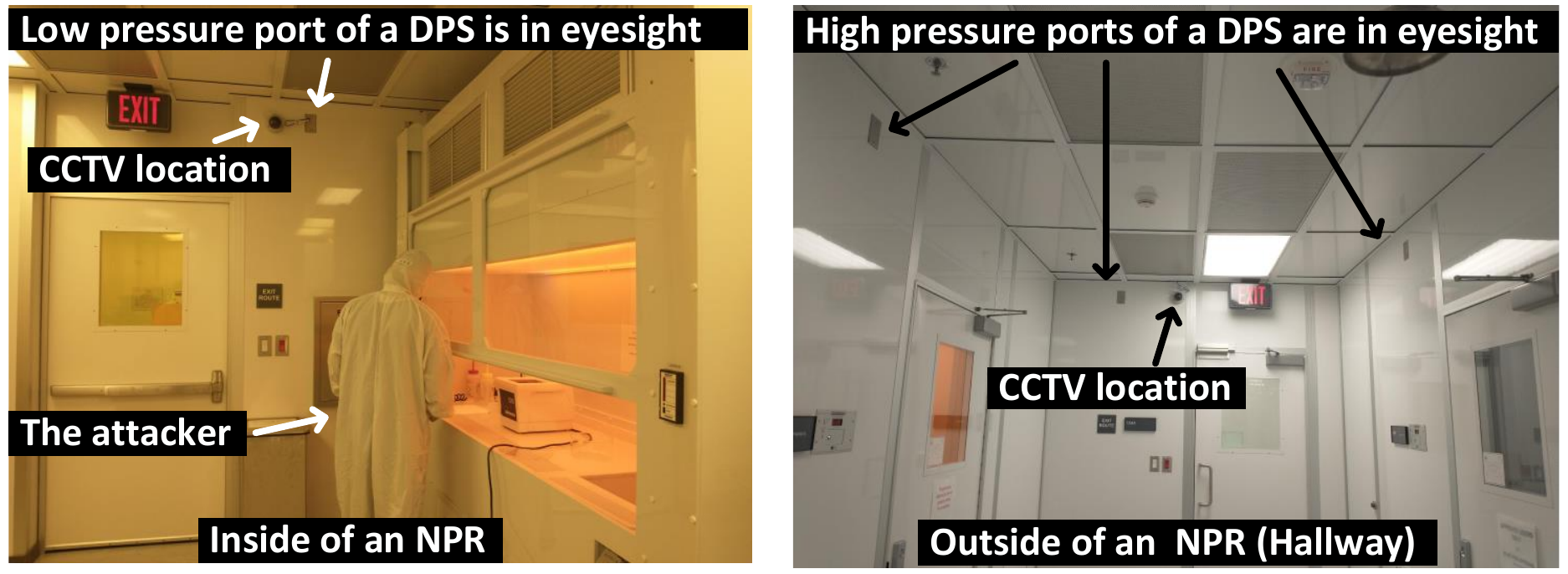}
\vspace{-1.20em}
\caption{Pressure ports of DPSs are in eyesight in NPRs.}
\label{fig:Expriement_Resonant_Frequency_pressure_port_cctv}
\vspace{-1.00em}
\end{figure}

\textbf{Attacker's capabilities:} The attacker can surreptitiously place an attack tool near the target pressure ports of a DPS used in an NPR. The attack tool has an \textit{audio source}. The audio source plays malicious music having a frequency equal to the resonant frequency of a DPS mounted in a target NPR. The audio source can be a simple cellphone or a speaker from an entertainment unit, such as televisions and radios, or CCTVs, placed in the vicinity of the pressure port of a target DPS. The low and high pressure ports are often mounted in eyesight, and placing the audio source near the target pressure port requires a \textit{brief one-time access}. Moreover, audio sources, such as televisions or CCTVs with speakers, are often installed in NPR facilities near the pressure ports (see Fig. \ref{fig:Expriement_Resonant_Frequency_pressure_port_cctv}). The audio source may have wireless controls allowing for remote communication. Therefore, the attacker can remotely control the timing of the attack and can pick a vulnerable time (e.g., infectious patient admitted in an NPR, ongoing bioresearch, etc.) for a maximal consequence. The authority of
the target NPR may not be aware of the attack model and
would possibly neglect the security implications of any audio source placed near the pressure ports in an NPR.

\textbf{Attacker's access level:} The access near the pressure port of a DPS needed for the attack can be possible in at least two scenarios. \textbf{First} (most likely), a malicious employee or a guest or a maintenance person, who has access to an NPR, may place the audio source near the pressure port. Though an NPR is restricted for unauthorized personnel, getting brief one-time access near the pressure port may not be difficult for an attacker in disguise of a guest or a maintenance person. \textbf{Second} is interdiction, which has been rumored to be used in the past \cite{spiegel,macri,snyder,chhetri2019tool} and has been recently proven to be feasible \cite{swierczynski2016interdiction}. During interdiction, a competitor can intercept the DPS during delivery or installation and may modify the DPS by placing an audio source inside and then proceed with delivery or installation to the NPR facility.

\textbf{Playing malicious music:} The attacker can play the malicious music in speakers to inject sound into DPS in the following three ways. \textbf{First}, the attacker can use a standard phishing attack to trick the authority into playing malicious music via email or a web page with autoplay audio enabled in CCTVs or televisions. \textbf{Second}, the attacker can play the malicious music using public radios. If some individuals place their radio near a pressure port, there is a good chance that the attack will be effective. \textbf{Third}, a physical proximity attack can happen if an attacker plays the music via a cell phone. 

\textbf{Outcomes of the attack:} The attacker changes the actual pressure reading of DPSs and fools the BMS to turn the negative room pressure into a positive pressure or reduce the negative pressure from a reference value. 
This will trigger an alarm and create chaos in the facility. Moreover, the NPR cannot work properly for what it is intended to design for and may not contain the deadly microbes. The intentional leak of deadly microbes from NPRs may result in bioterrorism. The potential for mass destruction by bioterrorism is evident from a report from the U.S. Office of Technology, which predicted that the release of 100 kg of anthrax spores in Washington, DC, would cause 130,000 to 3 million deaths, matching the lethal potential of a hydrogen bomb \cite{goldenberg2002early}. The CDC reviewed potential microbes, such as smallpox and viral hemorrhagic fever, as airborne bioweapons \cite{BARTLETT201284}. An intentional leak of these bioweapons from an NPR by an attacker can trigger a worldwide pandemic with a tremendous loss of human lives and monetary resources.



\textbf{Non-invasiveness:} The spoofing attack is non-invasive and is performed without making physical contact with the target DPS. The attacker don't need to directly access or physically touch the sensor readings. However, we expect that attackers can examine the behavior of a similar sensor subjected to acoustic impacts before launching an actual attack. 

\textbf{Attacker's resources and cost:} We assume that the attacker knows how the HVAC system works in NPRs and has a high school knowledge of resonance in DPSs. Moreover, a simple cell phone with a price of \$60 - \$100 can play the malicious music with a proper resonant frequency to attack the NPR.

\vspace{-00.100em}
\section{Threats in an NPR} 
\label{subsec:Threats in DPSs}


Here, we find the resonant frequency of DPSs used in NPRs by thorough experiments and explain how the resonance can be affected by different factors in an NPR.

\subsection{Sound wave as a threat to DPSs} 
\label{subsec:Sound Wave as a Stimulator to DPSs}

\,\,\,\,\,\,\,\textbf{Sound wave:} 
Sound is frequently referred to as a pressure wave since it is made up of a repeating pattern of high and low-pressure regions traveling across a medium \cite{Feynman}. 

\textbf{Threat to DPSs:} 
As a result, when sound waves collide with the diaphragms of DPSs, the diaphragm starts vibrating with the same frequency of sound. 
Therefore, having the above knowledge, a smart attacker can use a sound with a frequency equal to the resonant frequency of the DPS to create a \textit{resonance} and artificially displace the diaphragm in its maximal amplitude. The forged displacement of the diaphragm can change the pressure reading of a DPS by introducing overshooting in the actual pressure waveform. 

\subsection{Modeling sound effects on DPSs}
\label{subsec:Modeling Sound effects on DPSs}

We develop a model for how a sound wave perturbs the reading of a DPS. We measure the pressure as a linear combination of the original/equilibrium pressure $P_o(t)$ without a sound, and the external sound pressure $P_s(t)$. After a sound played at a frequency $f$, with an amplitude $A_0$, velocity $v$, and phase $\phi$ from a distance $d$, the total measured pressure $P(t)$ by a DPS can be modeled as:

\vspace{-0.7000em}
\begin{equation}
\begin{aligned}
P(t) &= P_o(t) + P_s(t)\\
    &= P_o(t) + h(d, f) \cdot A_0cos(2 \pi f t + d/v + \phi )
\vspace{-00.70em}
\label{eqn:shifitng}
\end{aligned}
\end{equation}

where $h(d, f)$ represents the attenuation of a sound wave, which depends on distance $d$ and frequency $f$ of the audio source. 
If the frequency $f$ of the sound wave is equal to the DPS's resonant frequency $f_r$, the impact $P_s(t)$ will be maximum for a target DPS.

It should be clear from the above explanation that the attacker, at first, needs to identify the resonant frequency $f_r$ of the DPS to orchestrate an attack. However, datasheets of the pressure sensors used in NPRs do not provide information related to their resonant frequencies. 
Therefore, we use thorough experiments to find the resonant frequency discussed in detail in the next sections. 


\begin{figure}[h!]
\vspace{-1.0em}
\centering
\includegraphics[width=0.28\textwidth,height=0.16\textheight]{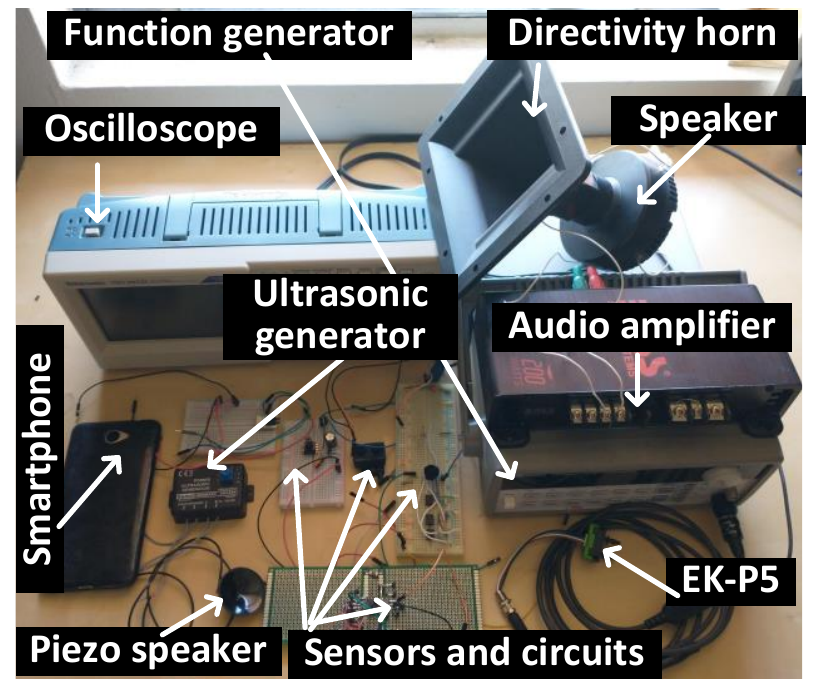}
\vspace{-1.1em}
\caption{Experiment setup for different DPSs.}
\label{fig:setup}
\vspace{-1.30em}
\end{figure}

\begin{table*}[ht!]
    \centering
    \caption{Summary of the resonant frequencies of Transducer Based Pressure Sensors (TBPSs) \textit{without} a sampling tube.}
    \vspace{-01.1800em}
    \begin{tabular}{p{0.2cm} | p{3.3cm}| p{1.7cm}|l|l|l|l|p{2.050cm}}
    \hline
        \cellcolor [gray]{0.85}\textbf{Sl.} & \cellcolor [gray]{0.85}\textbf{Sensor} & \cellcolor [gray]{0.85}\textbf{Manufac.} & \cellcolor [gray]{0.85} \textbf{Type} & \cellcolor [gray]{0.85} \textbf{Transducer} & \cellcolor [gray]{0.85} \textbf{Pressure range} & \cellcolor [gray]{0.85} \textbf{Interface} & \cellcolor [gray]{0.85} \textbf{Resonant freq.} \\ 
        \hline
        \hline
        1 & P1K-2-2X16PA \cite{P1K}  & Sensata  & Differential & Piezoresistive   & 0 to 500 Pa  & Analog & 790 - 800 Hz \\ 
        \hline
       2 &  MPVZ5004GW7U \cite{MPVZ500} & Freescale & Gauge  & Piezoresistive   & 0 to 3.92 kPa & Analog & 1750 - 1800 Hz \\
       \hline
        3 & SDP810-250PA \cite{SDPseries} & Sensirion  & Differential & Thermal mass-flow & $\pm$250 Pa  & Digital & 760 - 780 Hz \\ 
        \hline
        4 & SDP810-500PA \cite{SDPseries}  & Sensirion  & Differential & Thermal mass-flow & $\pm$500 Pa  & Digital & 870 - 890 Hz \\ 
        \hline
        5 & TBPDPNS100PGUCV \cite{TBPD} & Honeywell & Gauge  & Piezoresistive & 0 to 689 kPa & Analog & \cellcolor [gray]{0.85} not found \\ 
        \hline
        6 & P993-1B \cite{P9931b} & Sensata  & Differential & Capacitive & $\pm$248 Pa  & Analog & 740 - 750 Hz \\ 
        \hline
        7 & NSCSS015PDUNV \cite{NSCSS} & Honeywell & Differential & Piezoresistive & $\pm$103 kPa & Analog & \cellcolor [gray]{0.85} not found  \\ 
        \hline
        8 &  A1011-00 \cite{SeriesA1} & Sensocon & Differential & Piezoresistive & 0 to 60 Pa & Digital & 680 - 690 Hz \\
        \hline
    \end{tabular}
    \vspace{-1.20em}
    \label{table:SensorSpecAndResonantFreq}
\end{table*}

\vspace{-0.420em}
\subsection{Experimental setup}
\label{subsec:Identifying Resonant Frequencies}




Figure \ref{fig:setup} depicts the experimental setup to evaluate the resonant frequency of TBPSs. We produce a single-tone sound wave at different frequencies from an audible value of $50$ Hz to an inaudible value of $40$ kHz with the following three different audio sources. 

  
\text{\textbf{1. Source 1:}} We use a Samsung Galaxy S10 smartphone \cite{samsung_galaxy} to generate frequencies within $50$ Hz to $13$ kHz. We use an app named Function Generator to sweep frequencies within the specific frequency range using the smartphone, which has a \textit{sound pressure level (SPL)} of $\sim$ 80 dB \cite{samsung_galaxy_sound} at its maximum volume at 1-inch distance.

\vspace{0.1em}

\text{\textbf{2. Source 2:}} We use a function generator \cite{functiongenerator}, a 200 W audio amplifier (part\# BOSS Audio Systems R1002 \cite{audioamplifier}), a speaker (part\# Goldwood Sound Module \cite{soundspeaker}), and a directivity tweeter horn (part\# GT-1188 \cite{horn}) to generate frequencies within $100$ Hz to $18$ kHz. 
The directivity horn is connected with the speaker to direct the sound to the target sensor. This setup can generate an SPL up to $\sim$ 95 dB at 1-inch distance. The reason for using audio \textit{source 2} when we have the audio \textit{source 1} is to test the sensors with a higher SPL. We use an app named Sound Meter \cite{Soundmeter} to measure the SPL.

\vspace{0.1em}

\text{\textbf{3. Source 3:}} We use an ultrasound generator (part\# Kemo Electronic M048N \cite{ultrasonicsound}), a piezo speaker (part\# ToToT Ultrasonic Speaker \cite{piezospeaker}) to generate frequencies within a range of $15$ kHz to $40$ kHz. 
\vspace{0.1em}

We test 8 industry-used TBPSs from 5 different manufacturers including analog and digital types (see Table \ref{table:SensorSpecAndResonantFreq}). Out of the 8 sensors, 6 of them are DPSs, and 2 of them are gauge pressure sensors (see Section \ref{subsec:Transducer based pressure sensors (TBPSs)}). We use gauge sensors to identify that not only the DPSs but also the gauge pressure sensors have resonant frequencies that can be utilized by an attacker. This supports the idea that if an NPR uses a gauge pressure sensor instead of DPSs, an attacker can also target those NPRs. Therefore, our attack model will work for any TBPSs irrespective of gauge pressure sensors and DPSs.

The experimental setup is placed inside an acoustic isolation chamber to avoid external noise. To read and log the pressure measurements, we utilize an oscilloscope for analog TBPSs and a Ek-P5 \cite{datasheetEKP5} test kit connected with our laptop for digital DPSs.

Please note that a few pressure sensors require a separate unique circuit for testing, data collection, and signal conditioning. Therefore, we build a separate signal conditioning circuit for each of the sensors that requires it. As an example, a signal conditioning circuit using an instrumentation amplifier to collect data from a DPS with part\# NSCSSNN015PDUNV is shown in Appendix \ref{appendix:Signal_conditioning_circuit}.

\vspace{-0.3em}
\subsection{Evaluating the resonant frequency}
\label{subsec:Evaluation of the resonant frequenc}


A single tone sound having a frequency between $50$ Hz to $40$ kHz with a $10$ Hz increment is applied to \textit{one of the two ports} of a DPS or to a single port of a gauge pressure sensor in our testbed \textit{without} a sampling tube. We vary the frequency every 3 ms and record the data for every frequency using an oscilloscope for analog gauge/DPSs or using the Ek-P5 test kit for digital DPSs. We maintain the SPL within $\sim$ 35 - 95 dB from 2 cm in our experiments.

We examine the difference in the sensor readings with
and without sound signal. When there is no sound wave present, the two input ports of a DPS or a single input port of a gauge pressure sensor measure the unperturbed pressure from the environment. As a result, the intended output of the sensor should be zero in the absence of the single tone sound wave. When the single tone sound is applied to an input port of a DPS or a gauge sensor, the output of the target sensor starts oscillating. The oscillations reach a peak value at a resonant frequency of the target pressure sensor. 

\begin{figure}[h]
\vspace{-1.20em}
\centering
\includegraphics[width=0.48\textwidth,height=0.14\textheight]{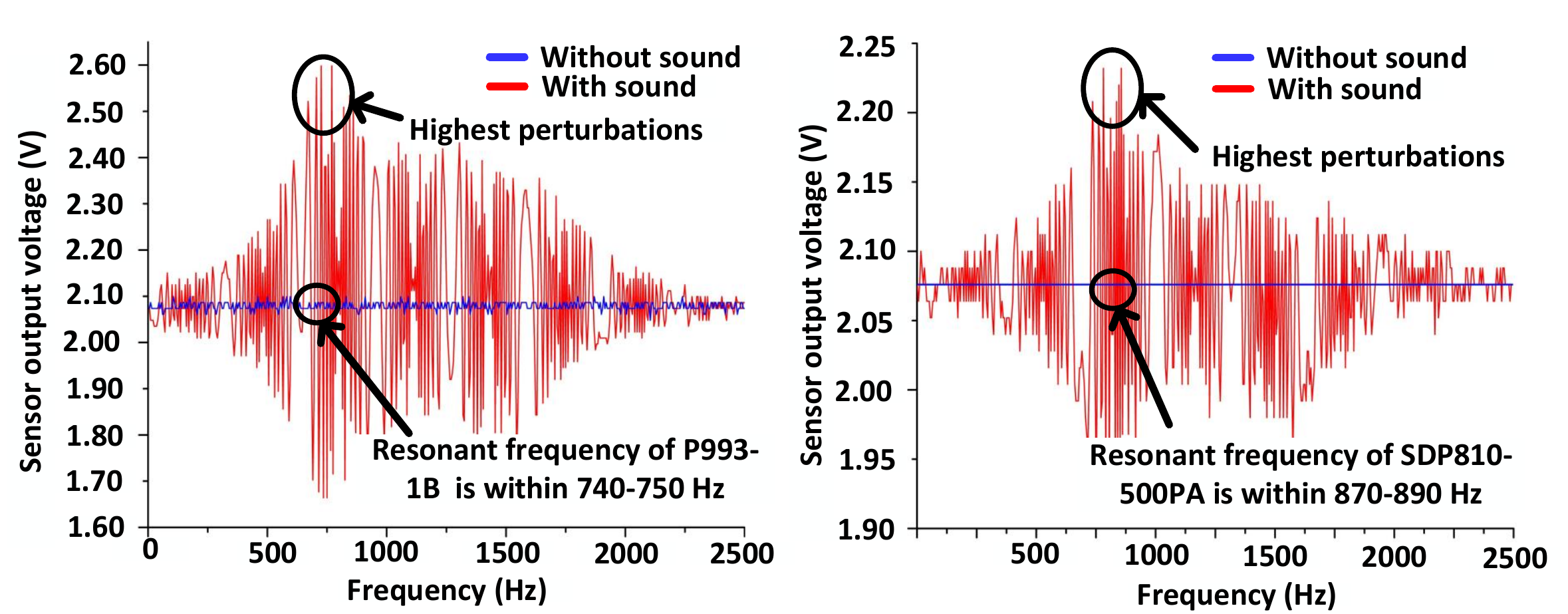}
\vspace{-2.20em}
\caption{Sound injection effect on (left) P993-1B and (right) SDP810-500PA pressure sensors for different frequencies.}
\label{fig:Expriement_Resonant_Frequency_anomo}
\vspace{-1.00em}
\end{figure}

Two examples are shown in Fig. \ref{fig:Expriement_Resonant_Frequency_anomo} as a proof-of-concept to support our observations on resonant frequencies. The outputs from an analog DPS with part\# P993-1 and a digital DPS with part\# SDP810-500Pa are shown in Fig. \ref{fig:Expriement_Resonant_Frequency_anomo} (left) and (right), respectively. The blue color is the sensor output before applying the sound, and the red color is the sensor output after applying the sound. It is clear from Fig. \ref{fig:Expriement_Resonant_Frequency_anomo} that the sensor output has the largest perturbations within 740-750 Hz for an analog DPS with the part\# P993-1B and within 870-890 Hz for a digital DPS with the part\# SDP810-500Pa.


Table \ref{table:SensorSpecAndResonantFreq} summarizes the experiment's findings on resonant frequencies. 
According to our findings, 6 of the 8 pressure sensors resonated in response to the applied sound wave. We find that the detected resonant frequencies range from $\sim$600 Hz to $\sim$1800 Hz, which are in the audible range.

We are unable to detect the resonant effect in 2 of the 8 sensors: part\# TBPDPNS100PGUCV and NSCSS015PDUNV. 
We observe from Table \ref{table:SensorSpecAndResonantFreq} that with the increase of the pressure range, the value of the resonant frequency increases. 
The reason behind this is that the sensors, which work in high pressure range, have more stiff diaphragms compared to those sensors, which work in low pressure range. For example, MPVZ5004GW7U has a higher resonant frequency than P1K-2-2X16PA because of its higher pressure range. 
Therefore, it is possible that the resonant frequencies of TBPDPNS100PGUCV and NSCSS015PDUNV may fall outside of 40 kHz, which is the highest test frequency we use in our experiments.

\vspace{-0.10em}
\subsection{Why resonant frequencies in audible range?}
\label{subsec:Why resonant frequencies are audible?}

An interesting observation from Table \ref{table:SensorSpecAndResonantFreq} is that all resonant frequencies of the DPSs used in NPRs fall in the audible range. We only experimented with 8 sensors used in NPRs. Can we conclude from our experiments that most of the sensors used in NPRs typically have resonant frequencies in the audible range? The answer is \textit{Yes}. 

\textbf{Reason}: Table \ref{table:countries_NP_values} shows that NPRs need to maintain a low negative pressure within 2.5 Pa to 15 Pa. Therefore, DPSs used in NPRs are selected to have high sensitivity in the low Pa range for an accurate measurement. The sensors working in the low pressure range have less stiff diaphragms compared to those sensors working in the high pressure range \cite{diaphragmstiffness}. Eqn. \ref{eqn:resonant_frequency} indicates that resonant frequency decreases in a square-root fashion with the decrease of stiffness of the diaphragms. Therefore, the DPSs working in a pressure range of few Pa, typically have less stiff diaphragms with low resonant frequencies typically in audible range (i.e., <20 kHz). 

\vspace{-0.20em}
\subsection{Factors influencing the resonant frequency}
\label{subsec:Factors not influencing the resonant frequencies}

We measure resonant frequencies in Table \ref{table:SensorSpecAndResonantFreq} by directly applying the sound wave to the input ports of a pressure sensor. However, sampling tubes and a pressure pick-up device are often connected with the pressure ports of a DPS (see Fig. \ref{fig:NPR_construction} and \ref{fig:DPS_pressure ports}) to pick up the pressure from a target location. 
The \textit{geometric properties} of the sampling tube affect the characteristics of the DPS's transducer systems. As a result, the resonant frequency of DPSs \textit{with} sampling tubes differs from the value \textit{without} sampling tubes. 

\begin{figure}[h]
\vspace{-1.20em}
\centering
\includegraphics[width=0.3\textwidth,height=0.09\textheight]{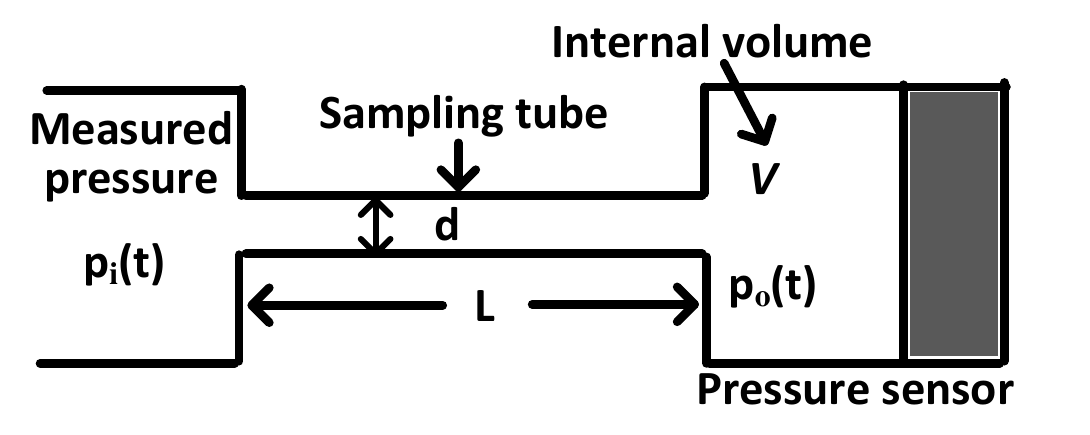}
\vspace{-1.420em}
\caption{Modeling sound pressure inside of a DPS having a sampling tube as a Helmholtz resonator.}
\label{fig:helmholtz_resonator}
\vspace{-1.300em}
\end{figure}

\textbf{Helmholtz resonators:} A pressure sensor with a sampling tube can be modeled as Fig. \ref{fig:helmholtz_resonator}. Let's denote the internal volume of the sensor by $V$, and the internal diameter and length of the tube by $d$ and $L$, respectively. As the sensor's internal volume and the connecting tube are similar to a structure having a cavity with a narrow neck, a pressure sensor with a tube is a basic form of discrete Helmholtz fluid resonator \cite{curry1990experimental,bajsic2007response}. The fluid in the tube acts as the oscillator mass, while the compressible fluid in the cavity acts as the oscillator spring. The Helmholtz resonator can be simplified by a second-order dynamic system (see Section \ref{subsec:Resonant frequency of a DPS and resonance}), which yields the following relation between the
sampling tube inlet pressure $p_i(t)$ and the sensor output pressure $p_o(t)$:

\vspace{-0.6000em}
\begin{equation}
\begin{aligned}
\dfrac{d^2p_o}{dt^2} + 2\xi \omega _h \dfrac{dp_o}{dt} + \omega_h ^2p_o = \omega _h ^2p_i
\vspace{-00.60em}
\label{eqn:helmhpltz_equation}
\end{aligned}
\end{equation}

where $\omega _h$ = 2$\pi f_h$, $f_h$ is the overall resonant frequency of the sensor with a tube, and $\xi$ is the damping ratio. The resonant frequency $f_h$ of the sensor with a tube can be expressed as:

\vspace{-0.5000em}
\begin{equation}
\begin{aligned}
f_h = \dfrac{1}{2 \pi} v\sqrt{\dfrac{A S}{L V M}} 
\vspace{-00.50em}
\label{eqn:overall_resonant_frequency}
\end{aligned}
\end{equation}

where $v$ is the sound velocity in air, $A$ is the internal cross-sectional area of the tube, $S$ is the stiffness of the diaphragm, $M$ is the mass of the pressure medium and diaphragm. Eqn. \ref{eqn:overall_resonant_frequency} indicates that the resonant frequency of a DPS with a tube increases with the increase of the tube's internal cross-sectional area $A$ and decreases with the increase of the tube length $L$. As the DPS used in NPRs has a standard diameter of its input ports, the diameter of the sampling tube is somewhat fixed. Therefore, we focus on the effect of sampling tube length on our attack model in the next section.

\vspace{-0.2000em}
\subsection{\textbf{Resonance with sampling tube in NPRs}} 
\label{subsec:Resonance with sampling tube in NPRs}

\begin{figure}[h]
\vspace{-1.20em}
\centering
\includegraphics[width=0.3\textwidth,height=0.14\textheight]{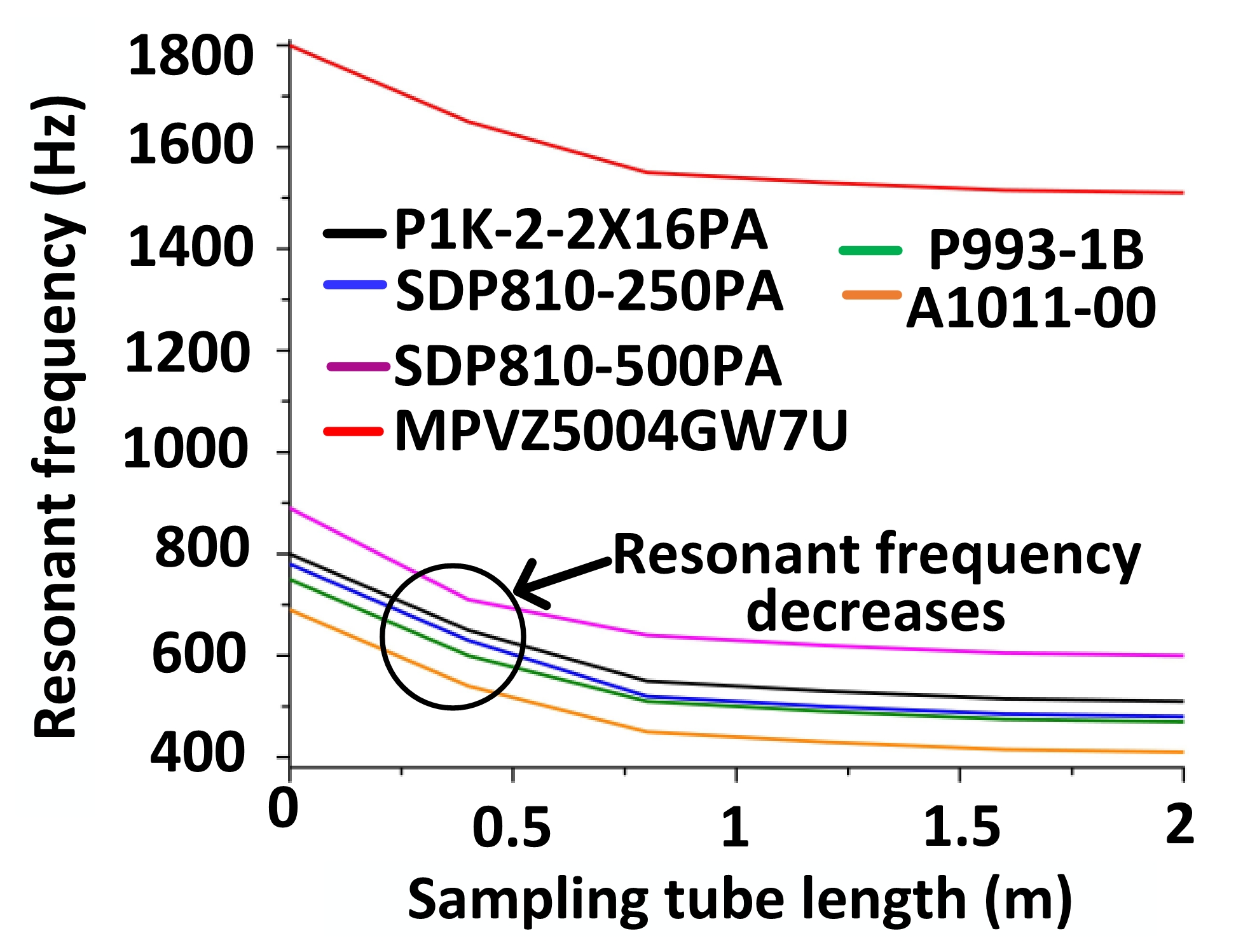}
\vspace{-1.320em}
\caption{Resonant frequency decreases with tube length.}
\label{fig:tube_length_resonant_frequency}
\vspace{-1.300em}
\end{figure}

Fig. \ref{fig:NPR_construction} and Fig. \ref{fig:DPS_pressure ports} show how the sampling tube is connected with the DPS's ports. For good sensitivity and error-free measurement, the DPS is placed close to the high and low pressure ports. Therefore, the length of the sampling tube is typically < 2 m. Therefore, we vary the length of the sampling tube up to 2 m with a 0.4 m increment for a diameter of 5/16 inch and calculate resonant frequencies for each of the 6 DPSs (i.e., having valid resonant frequency) from Table \ref{table:SensorSpecAndResonantFreq}. Fig. \ref{fig:tube_length_resonant_frequency} shows the results. We notice that with the increase of the sampling tube length, the sensor's overall resonant frequency $f_h$ reduces, supporting Eqn. \ref{eqn:overall_resonant_frequency}. 


\vspace{-0.400em}
\subsection{A wolf in sheep's clothing}
\label{subsec:A wolf in sheep's clothing - stealthiness}

It is evident from Section \ref{subsec:Resonance with sampling tube in NPRs} that the resonant frequencies of DPSs even with the sampling tube fall within audible range. Attacking DPSs with a sound just having resonant frequencies would make the attacker immediately identifiable because resonant frequencies will generate a "beep"-ish sound, raising a concern to the authority. 

We came up with a solution explained in \textbf{Section \ref{subsec:When HVAC and RPM use the same DPS}} to \textit{disguise} the resonant frequencies inside a popular music so that the attack will not be identifiable. Once the attacker injects the malicious music into DPSs, he/she can successfully create resonance in DPSs. This is referred to as putting \textit{"the wolf in sheep's clothing"} since it is the resonant frequency that has been disguised inside music. 

\section{Attacking a negative pressure room}
\label{sec:Attacking a negative pressure room}

As mentioned in Section \ref{subsec:How DPSs are deployed in an NPR}, the low pressure port of the DPS is exposed to the negative pressure room and the high pressure port is linked to a hallway, which is a reference space. If the pressure at the low pressure and high pressure port is denoted by $P_L$ and $P_H$, respectively, the DPS measures the differential pressure, $P_D$ as: 

\vspace{-01.000em}
\begin{equation}
\begin{aligned}
P_D = P_L - P_H
\vspace{-001.60em}
\label{eqn:differential_pressure}
\end{aligned}
\end{equation}

As mentioned in Section \ref{subsec:Construction of an NPR}, an NPR has an HVAC and an RPM system. There can be the following two scenarios depending on how the HVAC and RPM systems use the DPSs in NPRs. 

\textit{\textbf{First}}, the HVAC and RPM systems in NPRs use the \textit{same} DPS to control and monitor the negative pressure in an NPR. This scenario exists in modern facilities where both HVAC and RPM systems are automated and integrated with the BMS. 

\textit{\textbf{Second}}, the HVAC uses a DPS to maintain the negative pressure, and the RPM uses a \textit{separate} DPS to monitor the differential pressure in an NPR. 
Here, the RPM system only gives an alarm if the negative pressure falls below a threshold but is not responsible for maintaining a negative pressure in an NPR.

We discuss the above two scenarios below.

\subsection{When HVAC and RPM use the same DPS}
\label{subsec:When HVAC and RPM use the same DPS}

This scenario is easier for the attacker as he/she can attack both the HVAC and RPM systems of an NPR just by attacking a single DPS. The attacker can either inject sound to the low pressure port of the DPS if he/she is inside of the NPR and find that it is comparatively easier to access the low pressure port. Otherwise, the attacker can inject sound to the high pressure port of the DPS. 

\textbf{A simple resonance is not enough:} If the attacker creates resonance either by attacking the low pressure or high pressure port of the target DPS in the NPR, the resonance changes the original pressure reading by overshooting the original pressure level in both upward and downward directions (see Fig. \ref{fig:Expriement_Resonant_Frequency_anomo}). Therefore, the differential pressure reading $P_D$ in the DPS (Eqn. \ref{eqn:differential_pressure}) starts fluctuating. As a result, the \textit{supply fan} and the \textit{exhaust fan} immediately track the DPS's pressure fluctuations and vary their fan speed to maintain a static negative pressure inside of the NPR, following the algorithm \ref{alg:controlAlgorithm}. However, the rate of change in the pressure reading because of the resonance is high for a mechanical fan to track. Therefore, the supply fan and the exhaust fan cannot vary their speed with the high fluctuating rate. As a result, the negative pressure in the NPR only fluctuates a little bit and truly does not change on a large scale from the reference value. Moreover, the attacker does not have any adversarial control over it. Therefore, the attack can not induce any noticeable effect in the target NPR.

\begin{figure}[h]
\vspace{-00.680em}
\centering
\includegraphics[width=0.48\textwidth,height=0.21\textheight]{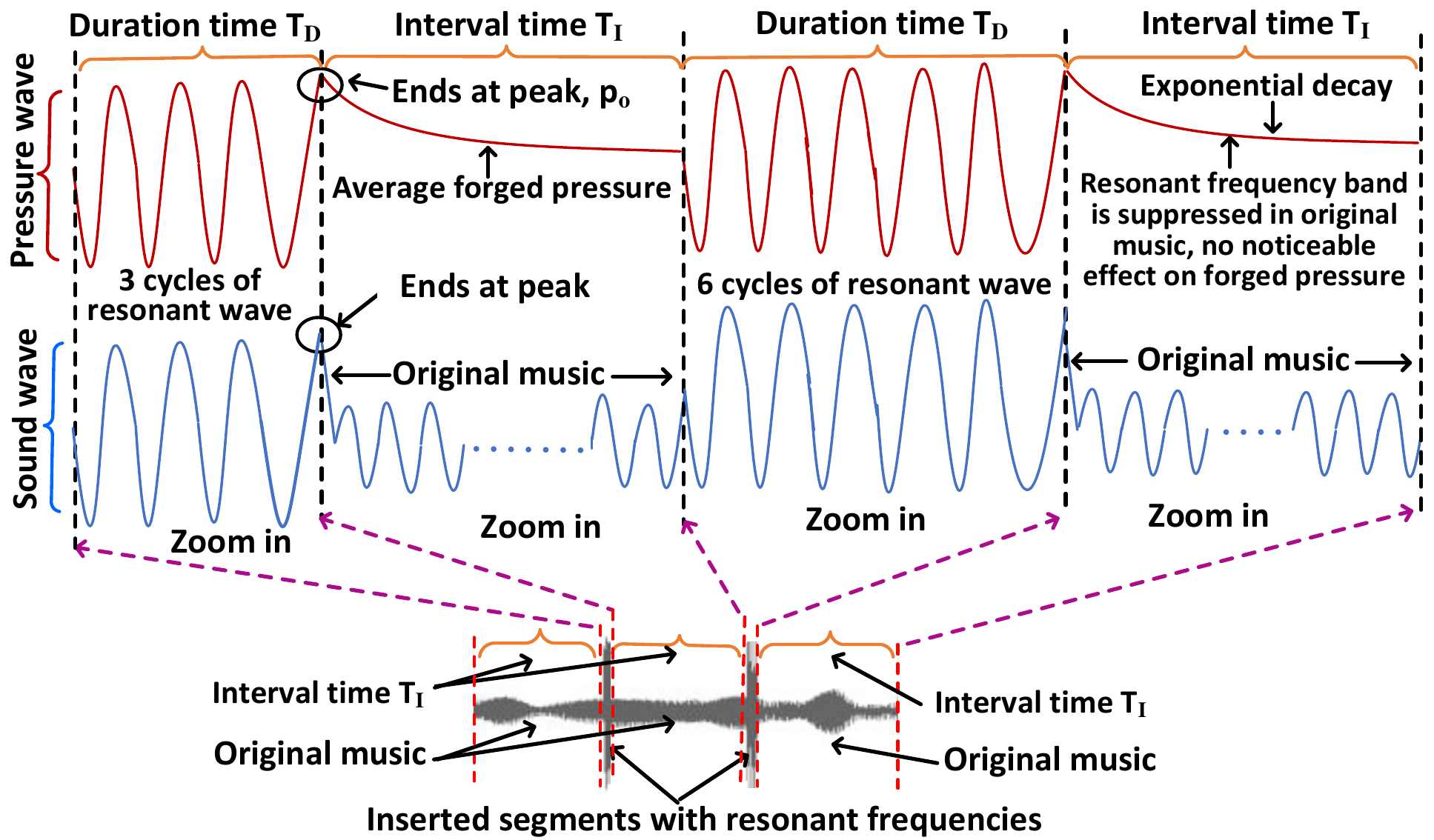}
\vspace{-2.30em}
\caption{Turning a popular music into an attack tool.}
\label{fig:wolf_in_sheeps_cloth}
\vspace{-0.00em}
\end{figure}

\textbf{A wolf in sheep's clothing:} To create a maximal change in the NPR's negative pressure, a smart strategy is adopted in addition to simply \textit{disguising} the resonant frequency band inside of music. The strategy is illustrated in Fig. \ref{fig:wolf_in_sheeps_cloth}. The resonant frequency is inserted into the music as a segment in a specific interval for a certain duration. Let us denote the interval by $T_I$ and duration by $T_D$. Every inserted segment of resonant frequency is \textit{ended at its peak} after duration $T_D$, and the same segment is inserted again in every interval $T_I$. When the inserted segment is ended at its peak, the corresponding pressure wave inside the DPS's transducer system also ends at its peak (see Fig. \ref{fig:wolf_in_sheeps_cloth}). As a DPS with a sampling tube is a second-order oscillating system (i.e., Helmholtz resonator), the pressure wave does not instantly fall to zero from the peak value. Instead, the pressure wave starts to attenuate from its peak exponentially following Eqn. \ref{eqn:critcally_damped} of a  damped $2^{nd}$ order system \cite{secondordersystem}.

\vspace{-0.6000em}
\begin{equation}
\begin{aligned}
p(t) = p_oe^{-\omega_ht} + (\omega_hp_o + v_o)te^{-\omega_ht}
\vspace{-001.10em}
\label{eqn:critcally_damped}
\end{aligned}
\end{equation}

where $p_o$ and $v_o$ are the initial pressure and velocity at peak, respectively, and $\omega_h$ is the angular resonant frequency. The term $v_o$ depends on the viscosity and density of the pressure medium.

The interval time $T_I$ is selected in such a way that the pressure wave never falls to zero. Therefore, there is always an \textit{average forged} pressure present inside the DPS's transducer system, originating from the injected music by the attacker. 
As the generated forged pressure has an \textit{average} value greater than zero and changes \textit{slowly}, the \textit{supply fan} and the \textit{exhaust fan} can track the pressure change in DPS, and they can vary their fan speed according to the pressure reading of the DPS. Therefore, this time the attack can induce a noticeable effect in the target NPR. 

\begin{figure}[h]
\vspace{-00.680em}
\centering
\includegraphics[width=0.48\textwidth,height=0.13\textheight]{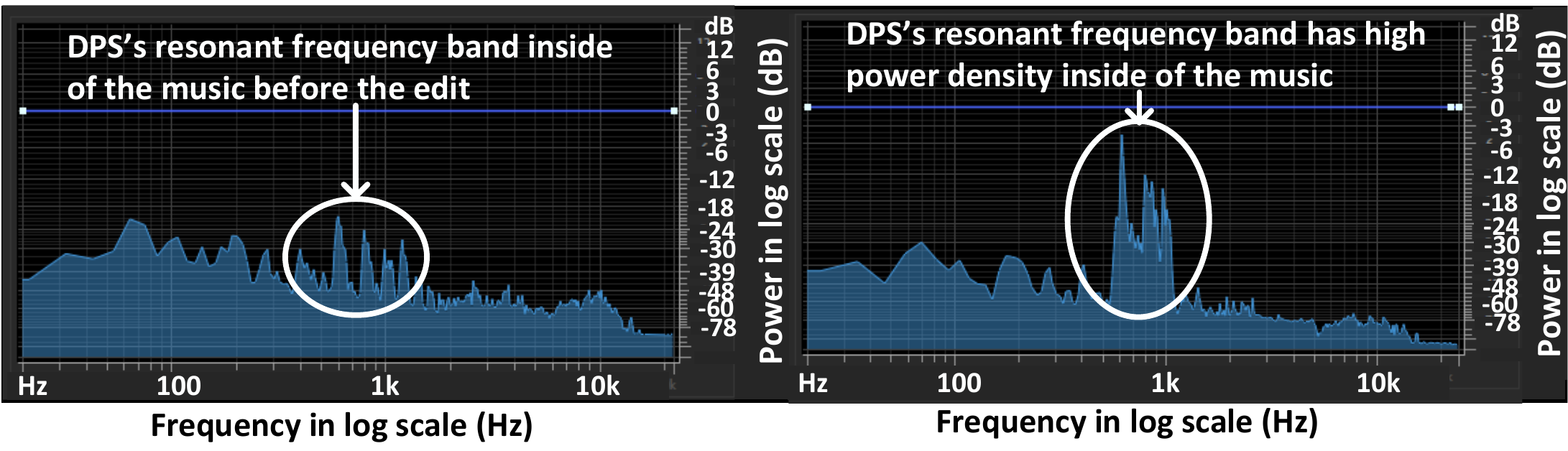}
\vspace{-2.30em}
\caption{High power density of resonant frequencies inside of a music because of the inserted segments.}
\label{fig:sound_signal_generation}
\vspace{-01.100em}
\end{figure}

Between two consecutive inserted segments of resonant frequency (i.e., in the interval time $T_I$), the original music is inserted by suppressing its resonant frequency components. Therefore, the original music does not have a noticeable effect on the forged pressure present in the interval $T_I$. Moreover, the inserted segment with the resonant frequency has $\sim$3.8x increased power density compared to the original music. Fig. \ref{fig:sound_signal_generation} shows this phenomena for SDP810-500PA, which has resonant frequency within 700 - 900 Hz (see Fig. \ref{fig:tube_length_resonant_frequency}). Therefore, the inserted segment can create a maximal effect in the NPR by turning a negative pressure into a positive one.

\textbf{Adversarial control:} The attacker can control the average forged pressure in the DPS's transducer system by controlling the interval time $T_I$ and duration time $T_D$.

The duration $T_D$ cannot be too small as a small $T_D$ cannot provide the inserted segment enough time to impact the DPS. The $T_D$ cannot be too large because the inserted segment with large $T_D$ can badly distort the music so that the attack might be identified. The duration of $T_D$ should be equal to or larger than the period of the resonant frequency so that at least one cycle of the resonant wave is accommodated inside of the duration $T_D$ (i.e., inserted segment).

\begin{figure}[h]
\vspace{-00.480em}
\centering
\includegraphics[width=0.4\textwidth,height=0.13\textheight]{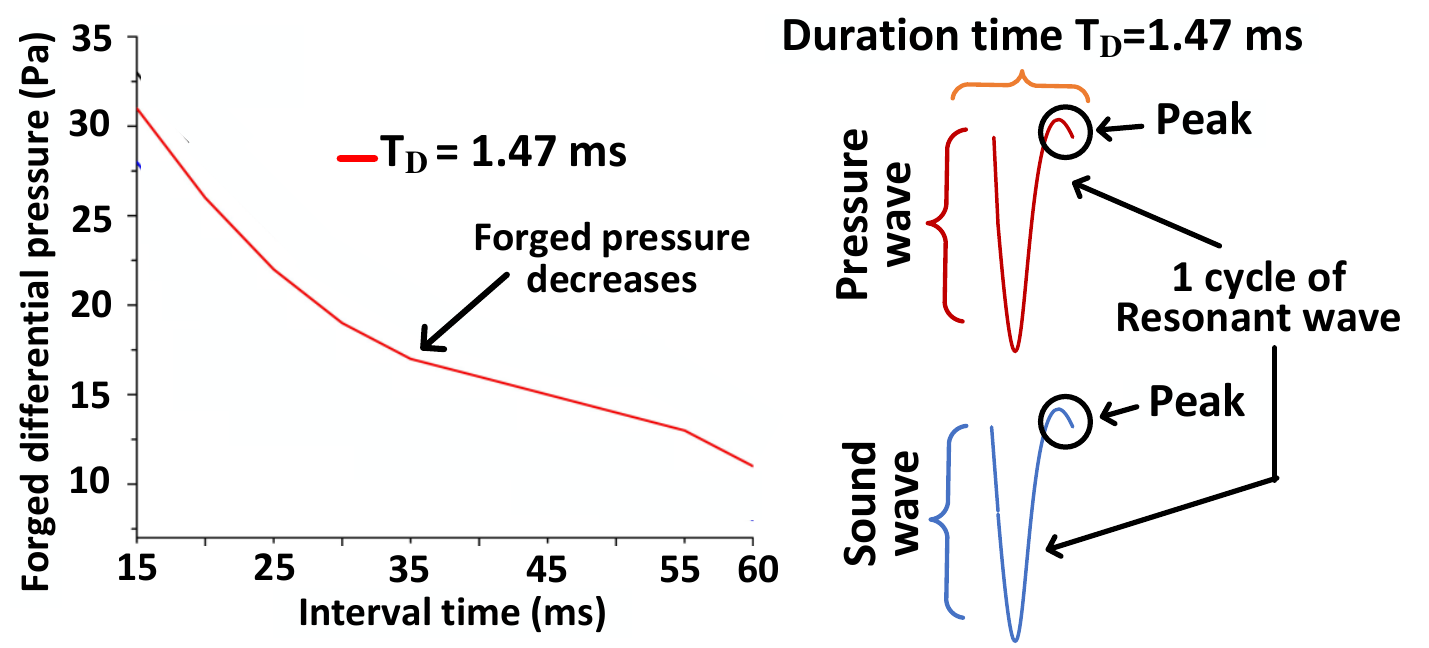}
\vspace{-1.20em}
\caption{Adversarial control using malicious music.}
\label{fig:adversarial}
\vspace{-1.000em}
\end{figure}

With a small interval $T_I$, the average forged pressure is increased. However, a small $T_I$ results in a large number of inserted segments that may distort the music significantly. We measured the forged pressure for a $T_I$ between 15 ms to 60 ms for a $\sim$65 dB sound for a DPS (part\# A1011-00) with a 1 m sampling tube. The sound is applied at 0.2 cm from the pressure port. The result is shown in Fig. \ref{fig:adversarial} for a duration time $T_D$ = 1.47 ms, which is equal to the period of the resonant wave of part\# A1011-00 (i.e., part\# A1011-00 has resonant frequency 680 Hz from Table \ref{table:SensorSpecAndResonantFreq}; 1/680 Hz = 1.47 ms).

As mentioned earlier, the resonant frequency can vary within a range depending upon the sampling tube length. As the attacker may not know the exact length of the sampling tube, the attacker may need to vary the duration time $T_D$ within a range to accommodate at least one cycle of the variable resonant wave for a maximal impact (Fig. \ref{fig:adversarial}). The attacker can also vary the number of cycles in the duration $T_D$ from one inserted segment to another inserted segment. For example, the first inserted segment in Fig. \ref{fig:wolf_in_sheeps_cloth} has 3 cycles, whereas the second segment has 6 cycles within the duration $T_D$.

\textbf{Tools for a malicious music:} The attacker selects  music and inserts segments of resonant frequencies within the music in a way already explained in Section \ref{subsec:When HVAC and RPM use the same DPS} using a software named \textit{Adobe Audition}. Though someone who has listened to the music many times before may identify the change in the music, the vast majority of people will either be oblivious of the change or will incorrectly ascribe the change in the music to a speaker issue. For example, we pick a popular song \textit{Hello} by \textit{Adele} and convert it into a malicious music in a way explained in Section \ref{subsec:When HVAC and RPM use the same DPS} for $T_D$ = 2 ms and $T_I$ = 15 ms. The malicious music is uploaded in the following link: {\color{blue}\url{https://sites.google.com/view/awolfinsheepsclothing/home}}

\textbf{Injecting music into the low pressure port:} Let us give an example to elaborate on the result of injecting music into the low pressure port. Suppose, before an attack, the pressure at a low pressure port $P_L$ = 10 Pa and at a high pressure port $P_H$ = 12.5 Pa. Therefore, the differential pressure from Eqn. \ref{eqn:differential_pressure} is $P_D$ = 10 - 12.5 = - 2.5 Pa, which is the reference differential pressure in the NPR. Suppose the forged pressure resulting from the injected malicious music into the low pressure port is 8 Pa. Now, after the attack, $P_D$ = (10 + 8 = 18) - 12.5 = 5.5 Pa. Therefore, the HVAC system will reduce the NPR's pressure from 18 Pa to 10 Pa to keep the differential pressure at -2.5 Pa. The reduction of 8 Pa in the NPR will result in a \textit{true} differential pressure of $P_D$ = (10 - 8 = 2) - 12.5 = -10 Pa. The injection of music into the low pressure port results in more negative differential pressure (i.e., - 2.5 Pa to - 10 Pa), which is actually good for keeping deadly microbes in the NPR. However, the abnormal change in negative pressure may trigger an alarm by the RPM system and create chaos in the facility. An attacker can use this chaos to initiate a stronger attack, such as stealing deadly microbes from biosafety cabinets as he is already inside of the NPR.

\textbf{Injecting music into the high pressure port:} Let us use the previous example to elaborate on the effect of injecting music into the high pressure port. If the forged pressure resulting from the injected music into the high pressure port is 8 Pa, the $P_D$ after the attack is 10 - (12.5 + 8) = -10.5 Pa. Therefore, the HVAC system will increase the NPR's pressure from 10 Pa to 18 Pa to keep the differential pressure at -2.5 Pa. The increase of 8 Pa in the NPR will result in a \textit{true} differential pressure of $P_D$ = (10 + 8 = 18) - 12.5 = 5.5 Pa, which is positive. The consequences of turning a negative pressure into a positive one in an NPR can be catastrophic as the NPR cannot contain the deadly microbes anymore, causing a potential leak of microbes from the compromised NPR. Moreover, an abnormal change in the negative pressure may trigger an alarm by the RPM system and create chaos in the facility. An attacker can use this chaos to initiate a stronger attack, such as entering the NPR and stealing deadly microbes from the biosafety cabinets.

\vspace{-0.32em}
\subsection{When HVAC and RPM use separate DPSs}
\label{subsec:When HVAC and RPM use different DPSs}

When the HVAC and RPM systems use separate DPSs, and if the attacker has a \textit{single} audio source, he/she should attack the high or low pressure port of the DPS connected with the HVAC system to change the negative pressure in an NPR. Because the HVAC system maintains the negative pressure in an NPR. However, if the attacker attacks the high or low pressure port of the DPS connected with the RPM system, only an alarm may be triggered, and chaos will be created in the facility, but it will not change the NPR's pressure. The attacker can use the attack model already explained in Section \ref{subsec:When HVAC and RPM use the same DPS} either to attack the HVAC or RPM system of an NPR. 


\textbf{A stronger attacker:} Suppose we consider a stronger attacker, who can use \textit{multiple} audio sources to attack the RPM and HVAC systems simultaneously. In that case, he/she can avoid the alarm triggered by the RPM system in the following way. 

Let us explain this attack model using the same example from Section \ref{subsec:When HVAC and RPM use the same DPS}. Let us assume the attacker injects the same \textit{forged} pressure of 8 Pa by music to the high pressure port of the DPS connected with the HVAC system. Therefore, the HVAC system similarly will increase the NPR's pressure from 10 Pa to 18 Pa, resulting in a positive differential pressure of 5.5 Pa. The RPM system will trigger an alarm for this abnormal change in the NPR's pressure. To prevent the alarm from being triggered, the attacker must need to inject the same 8 Pa \textit{forged} pressure to the high pressure port of the DPS connected with the RPM system. As a result, the RPM will measure differential pressure of 18 - (12.5 + 8) = -2.5 Pa, which is equal to the NPR's reference pressure. Therefore, the RPM system will not trigger any alarm, and the attack will remain unidentified, resulting in a stronger attack model.

However, if both of the high pressure ports of the RPM and HVAC systems are in \textit{close proximity}, the attacker can use a \textit{single} audio source to attack the NPR without triggering the alarm.

\subsection{Attacking multiple NPRs simultaneously}
\label{subsec:Attacking multiple NPRs simultaneously}

\begin{figure}[h!]
\centering
\includegraphics[width=0.45\textwidth,height=0.2\textheight]{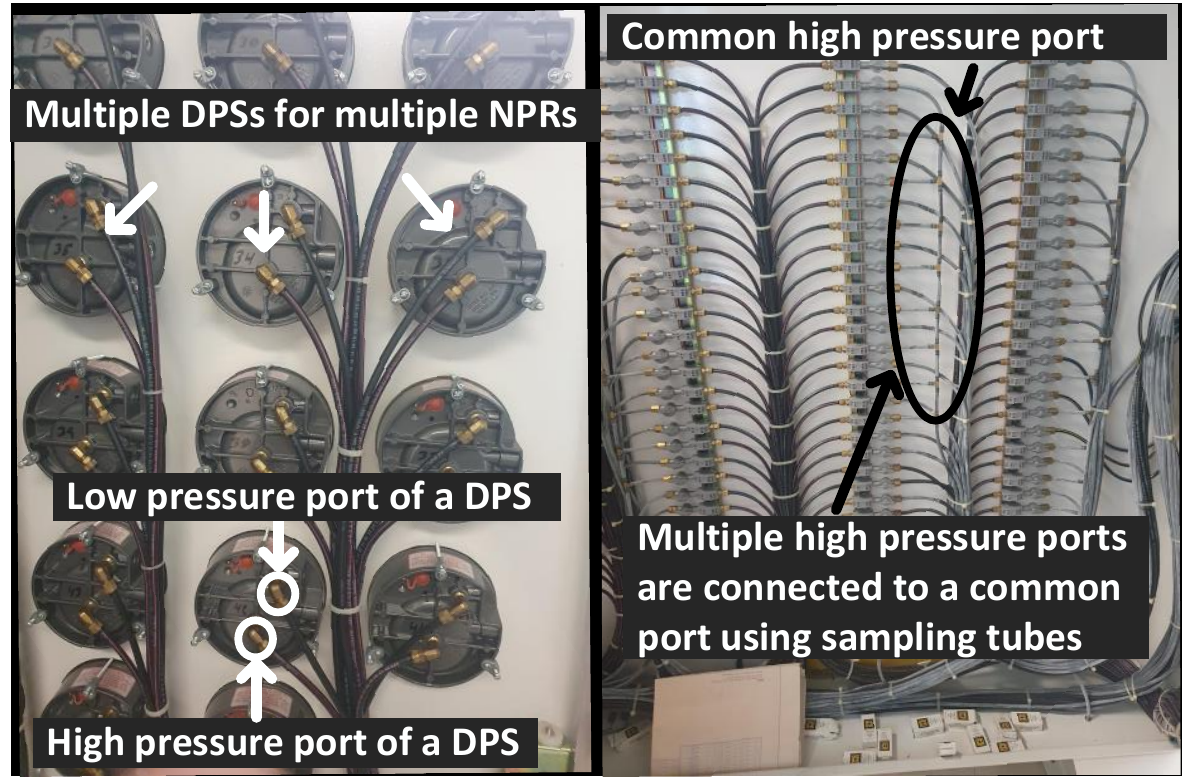}
\vspace{-1.0em}
\caption{Multiple high pressure ports are connected together to a common high pressure port.}
\label{fig:common_high_pressure_port}
\vspace{-01.610em}
\end{figure}

It is possible to simultaneously attack multiple NPRs just by injecting music into \textit{a single} high pressure port of the DPS connected with the HVAC or RPM systems. As we mentioned earlier, the high pressure port is located in hallway to measure the reference pressure, and the NPR maintains a negative pressure inside with respect to the reference pressure. If there are multiple NPRs in a facility and if all the NPRs use a common place (e.g., hallway) as their reference pressure, it is a common practice to connect all the high pressure ports from all the NPRs into \textit{one common} high pressure port to reduce cost. This is shown in Fig. \ref{fig:common_high_pressure_port}. As multiple NPRs share a common high pressure port, the attacker can simultaneously attack multiple NPRs just by attacking the common high pressure port in the facility. It can trigger a combined leak of deadly microbes from multiple NPRs and can create chaos in different parts of the facility. 

\vspace{-0.870em}


\section{Attack model demonstration}
\label{sec:Attack model demonstration}

\begin{figure}[h!]
\centering
\includegraphics[width=0.46\textwidth,height=0.2\textheight]{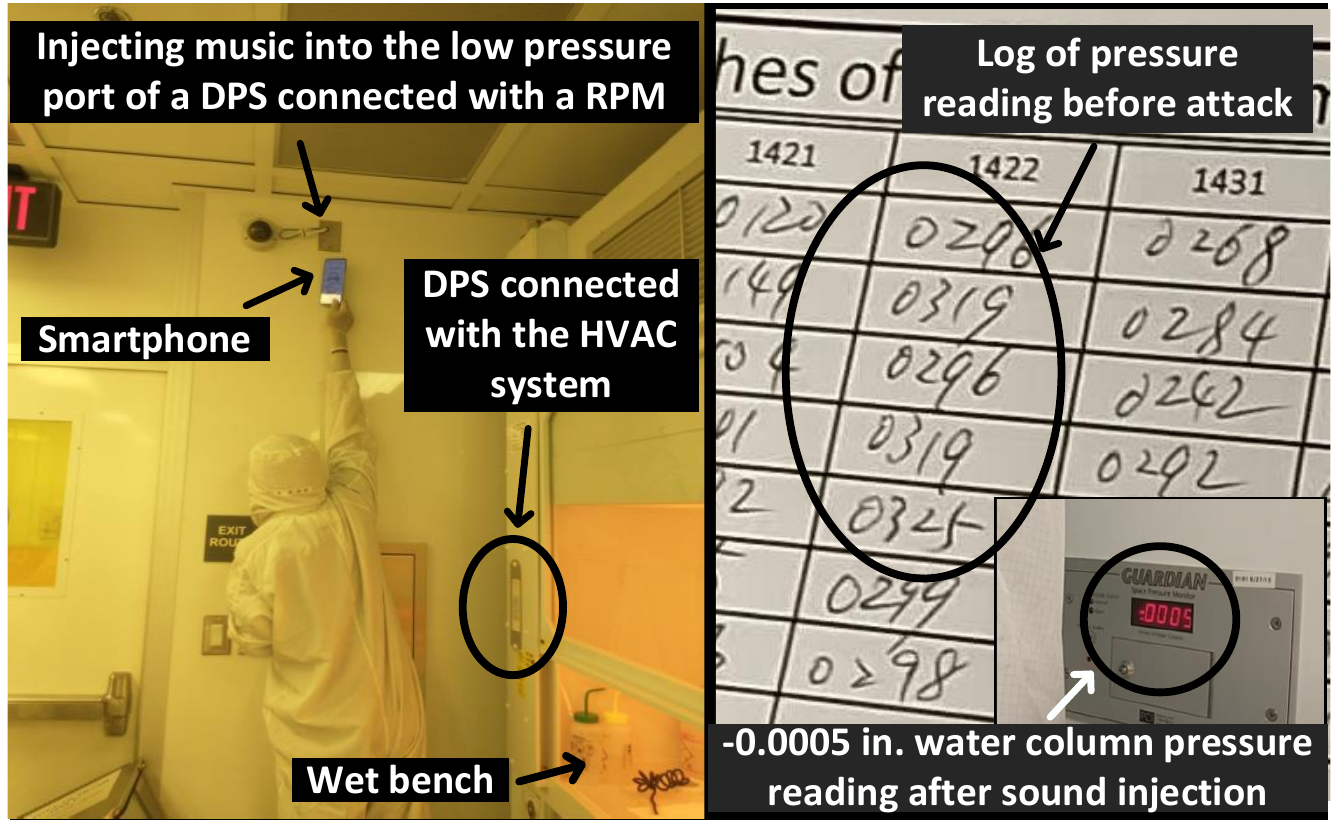}
\vspace{-1.1190em}
\caption{Attacking a practical NPR in a bioresearch facility.}
\label{fig:attack_model_demonstration}
\vspace{-01.0em}
\end{figure}

We demonstrate our attack at an FDA-approved NPR located in an anonymous bioresearch facility. The demonstration is shown in Fig. \ref{fig:attack_model_demonstration}. This facility uses separate DPSs for the HVAC and RPM systems. The location of the DPS connected with the RPM system is close to the exit door. The DPS connected with the HVAC is at the sidewall of the wet bench. The wet bench stores sensitive particles inside of it under negative pressure. The authority did not permit us to attack the DPS connected with the HVAC system due to safety protocols. Therefore, we only demonstrate the attack on the DPS connected with the RPM system.

\vspace{-0.760em}

We use a Samsung Galaxy S10 smartphone from a 0.1 cm distance with an SPL of $\sim$ 65 dB to inject the malicious music into the low pressure port of the RPM system for a room \#1422. We check the differential pressure for room \#1422 before the attack from a logbook. We can see that the negative pressure stays within a range of 0.0278 - 0.0325 inch water column (i.e., 6.9 - 8 Pa). After injecting music from the smartphone, the negative pressure reading in the RPM system changes to a positive pressure of  0.0005 inch water column (i.e.,  0.12 Pa). That is a 7 - 8 Pa pressure reading change in the RPM system due to an attack. A video demonstrating the attack model in the NPR is posted at the following link: {\color{blue}\url{https://sites.google.com/view/awolfinsheepsclothing/home}}

Though we are not permitted to attack the DPS connected with the HVAC system, according to the authority, our attack on the DPS connected with the HVAC system would create the same pressure change in the NPR.


\section{Attack model evaluation}
\label{sec:Attack model evaluation}

We already evaluate resonant frequencies of DPSs in Section \ref{subsec:Threats in DPSs} in detail. Here, we evaluate our attack model further for other parameters related to the DPSs and NPRs.

\begin{figure}[h!]
\vspace{-0.60em}
\centering
\includegraphics[width=0.38\textwidth,height=0.14\textheight]{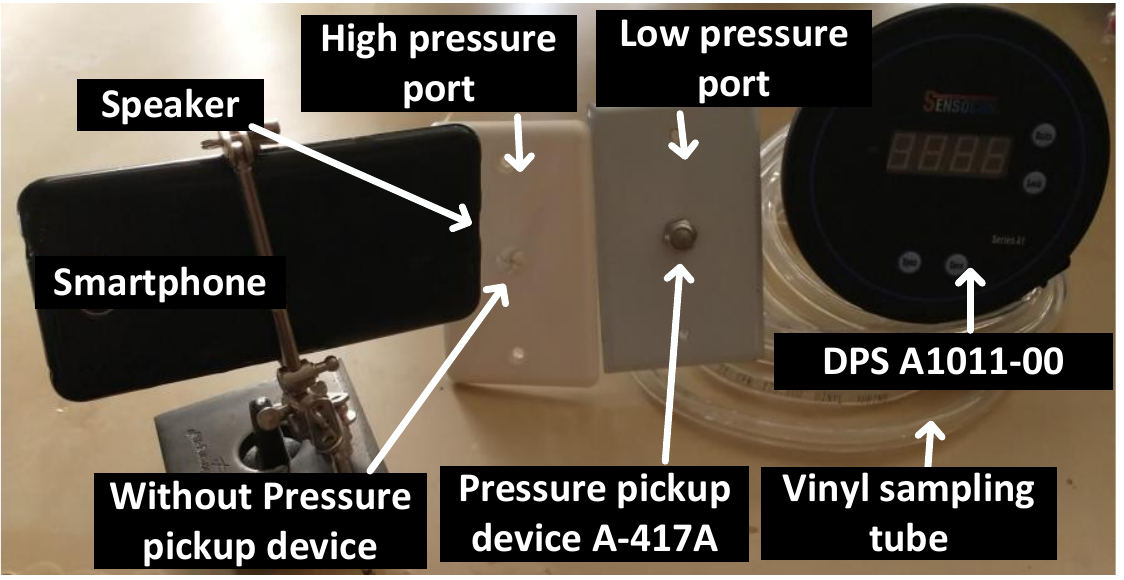}
\vspace{-1.0em}
\caption{Experimental setup for evaluating attack model.}
\label{fig:experimental_setup_simple}
\vspace{-01.410em}
\end{figure}

\subsection{Experimental setup}
\label{subsec:Experimental setup}

We already show our attack at an FDA-approved NPR in a bioresearch facility in Section \ref{sec:Attack model demonstration}. As it is not \textit{permitted} to vary different parameters of the DPS's transducer system located in the bioresearch facility, we prepare a testbed to evaluate our attack model. We use an industry used DPS from Sensocon with part\# A1011-00 \cite{Sensocon}, two vinyl sampling tubes having inner diameters of 3/16" and 5/16" \cite{vinyltube}, a pressure pickup device with part\# A-417A \cite{pressurepickup} and an oscilloscope in the testbed (see Fig. \ref{fig:experimental_setup_simple}). 

\vspace{-0.30em}
\subsection{Varying the tube length and diameter}
\label{subsec:Varying the sampling tube length}

\begin{figure}[h!]
\centering
\includegraphics[width=0.48\textwidth,height=0.14\textheight]{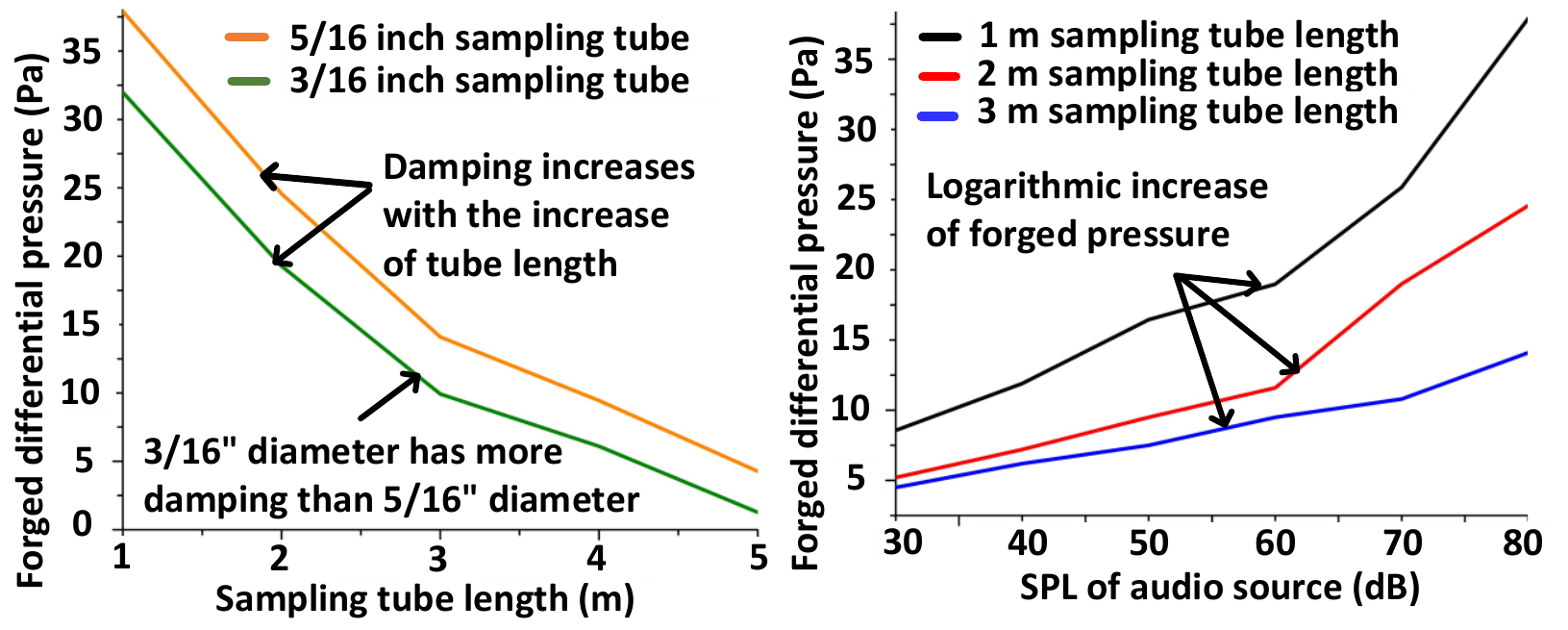}
\vspace{-2.260em}
\caption{(left) Impact of sampling tube length and diameter. (right) Impact of the SPL of the audio source on the attack }
\label{fig:sampling_tube_length_diameter}
\vspace{0.30em}
\end{figure}

We vary the sampling tube length from 1 m to 5 m with a 1 m increment for two inner diameters of 3/16" and 5/16". We connect the sampling tube and pressure pickup device with the input ports of the A1011-00 sensor and inject sound into one of the pressure ports with the Samsung Galaxy S10 smartphone from a 0.1 cm distance. The result is shown in Fig. \ref{fig:sampling_tube_length_diameter} (left). With the increase of the sampling tube length and the decrease of the sampling tube inner diameter, the sound damping inside the tube increases. Therefore, the forged differential pressure originated from the injected music reduces for larger length and smaller diameter.

\vspace{-0.0em}

\vspace{-0.0em}
\subsection{Varying the SPL of the audio source}
\label{subsec:Varying the SPL of the audio source}

A logarithmic scale known as Sound Pressure Level (SPL) is used to measure the loudness of a sound. SPL is measured in decibels (dB). We vary the SPL of the audio source (i.e., Samsung Galaxy S10) from 30 dB to 80 dB with a 10 dB increment for 1m, 2 m, and 3 m of sampling tube (5/16" diameter) lengths for a 0.1 cm distance from the pressure pickup device. The result is shown in Fig. \ref{fig:sampling_tube_length_diameter} (right). As with the increase of the SPL, the sound pressure from the audio source logarithmically increases. Therefore, the forged differential pressure also increases logarithmically. As sound damping increases with the increase of sampling tube length, the shorter sampling tube causes higher forged differential pressure.


\vspace{-0.50em}
\subsection{Varying the distance of the audio source}
\label{subsec:Varying the distance of the audio source}

\begin{figure}[h!]
\vspace{-01.20em}
\centering
\includegraphics[width=0.47\textwidth,height=0.15\textheight]{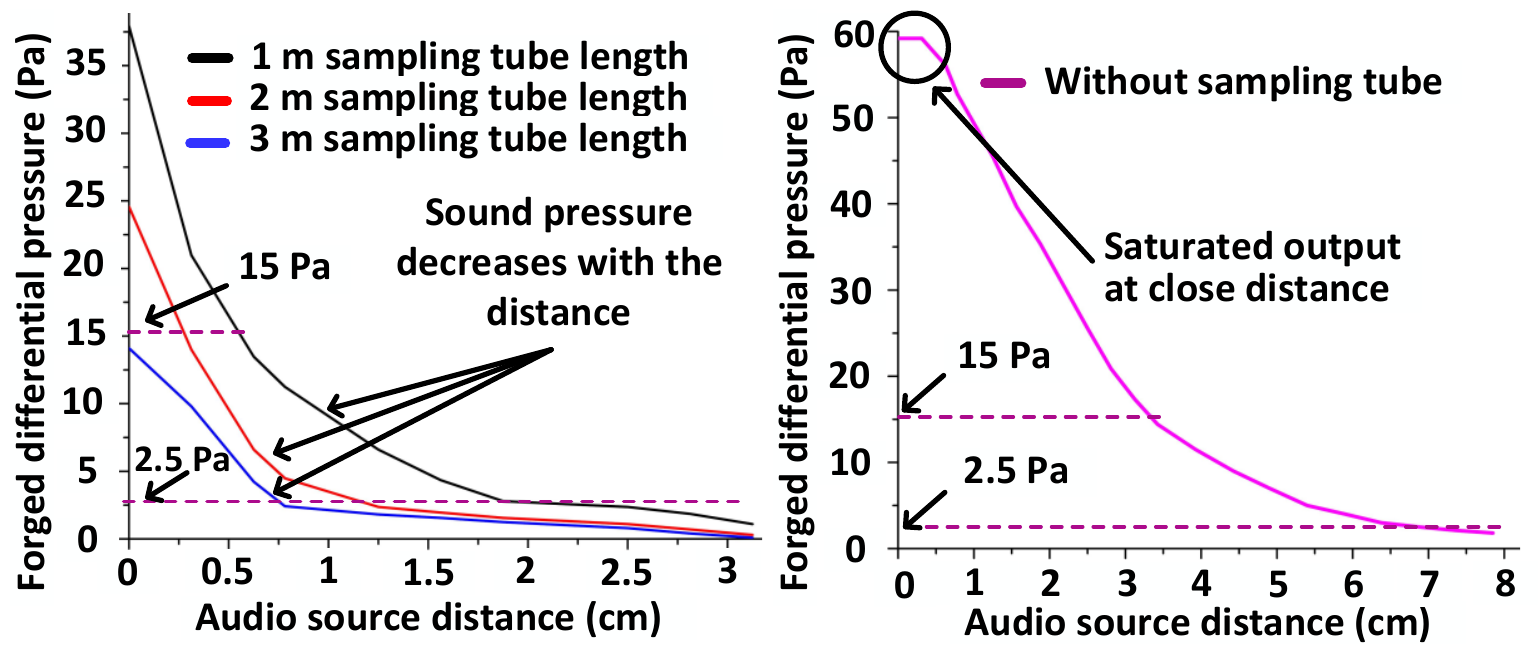}
\vspace{-1.20em}
\caption{Impact of audio source distance on the attack.}
\label{fig:audio_source_distance_tube_length}
\vspace{0.00em}
\end{figure}

We vary the distance of the audio source (i.e., Samsung Galaxy S10) from the pressure pickup device for 0 m (no sampling tube), 1 m, 2 m, and 3 m of sampling tube (5/16" diameter) lengths. The result is shown in Fig. \ref{fig:audio_source_distance_tube_length}. In acoustics, the SPL of a sound wave radiating from a point source decreases as the distance increases following the inverse-proportional law \cite{soundpropagation}: $SPL \propto 1/distance$. Therefore, the forged differential pressure also decreases with the increase of audio source distance from the pressure pickup device. Fig. \ref{fig:audio_source_distance_tube_length} (right) shows that an audio source has more impact on the DPS without a sampling tube (i.e., no dampening) with a saturated output.

\vspace{-1.900em}
\subsection{With and without a pressure pickup device}
\label{subsec:With and without pressure pickup device}

\begin{figure}[h!]
\vspace{-0.50em}
\centering
\includegraphics[width=0.37\textwidth,height=0.15\textheight]{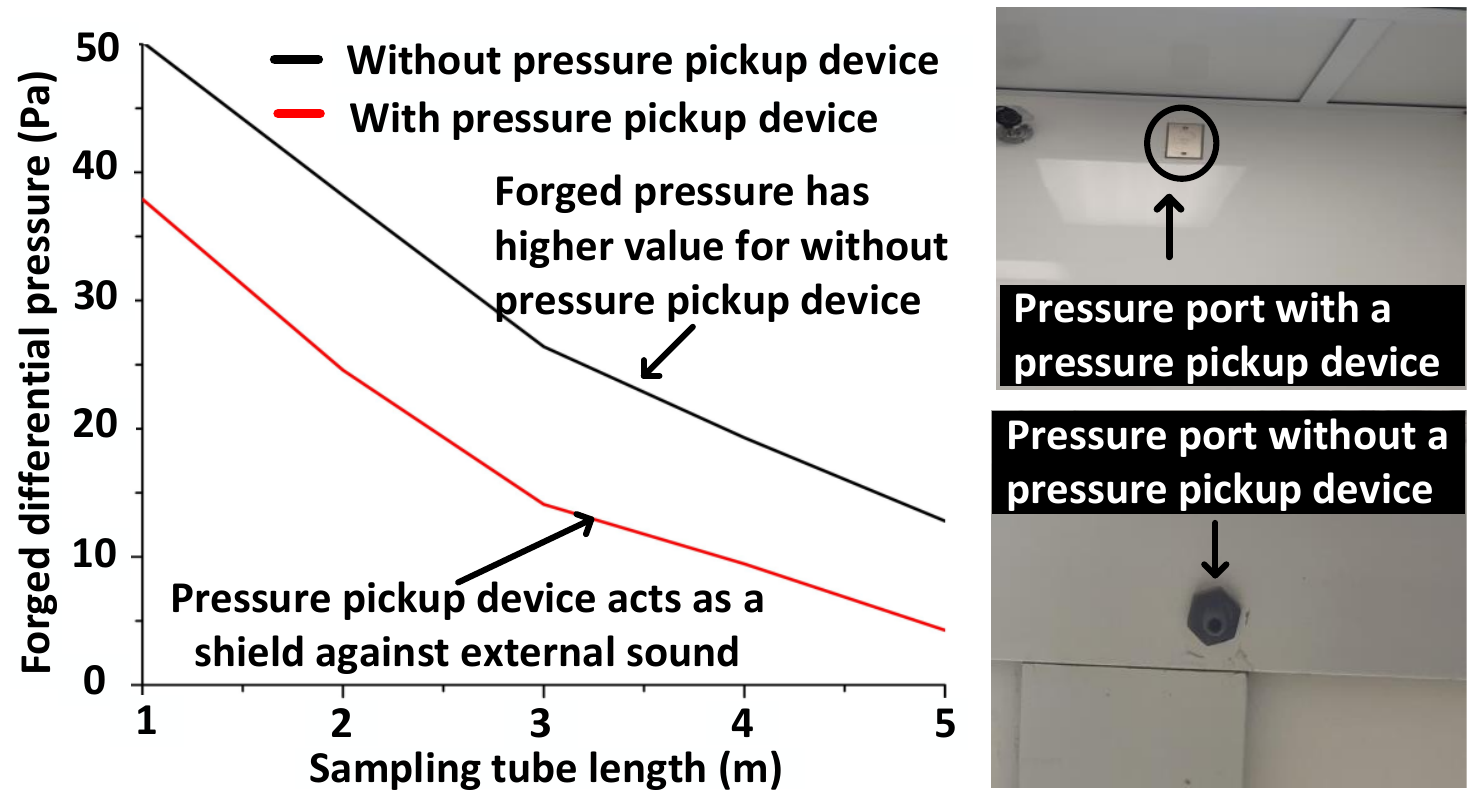}
\vspace{-1.0em}
\caption{Impact of the pressure pickup device on the attack.}
\label{fig:pressure_pickup_device}
\vspace{0.40em}
\end{figure}

A pressure pickup device is connected with the other end of the sampling tube and installed at the high and low pressure ports, mounted on the wall. The pressure pickup device increases the exposed area of the sampling tube end. Therefore, a small change in pressure can be sensed without an error. It is possible that some NPRs don't use pressure pickup devices; instead, a simple hole is mounted at the pressure ports. To evaluate the effect of the pressure pickup device, we inject music from a 0.1 cm distance into the pressure port with and without the pressure pickup device and vary the sampling tube length from 1 m to 5 m with a 1 m increment. We see from the results in Fig. \ref{fig:pressure_pickup_device} that the forged pressure is lower with a pressure pickup device. Because a pressure pickup device has foam gasket inside, which dampens the injected sound into it.

\section{Feasibility of the Attack}
\label{sec:Feasibility of the Attack}


\textbf{1. Audio source distance:} Section \ref{sec:Attack model evaluation} indicates that the sampling tube's length and audio source's distance 
can restrict the effectiveness of the attack. Moreover, Table \ref{table:countries_NP_values} indicates that the negative pressure has to be maintained between -2.5 Pa to -15 Pa for country-specific requirements. Now, Fig. \ref{fig:audio_source_distance_tube_length} (left) indicates that the audio source should be less than 0.6 cm away (1 m tube length) from pressure ports to generate a 15 Pa forged pressure, which can turn a -15 Pa negative pressure into a positive pressure. Fig. \ref{fig:audio_source_distance_tube_length} (left) also indicates that the audio source should be less than 2.5 cm away from the pressure port to generate a 2.5 Pa forged pressure, which can turn a -2.5 Pa negative pressure into a positive pressure. \textit{This indicates that the \textit{CDC} guidelines (i.e., -2.5 Pa) in Table \ref{table:countries_NP_values} can be \textit{impacted from a larger audio source distance} compared to the guidelines adopted in Taiwan and Australia.} 

However, the audio source needs to be in close proximity to the pressure ports to have a feasible attack. CCTV's with speakers and entertainment units are often located in such close proximity to the pressure ports. Moreover, Fig. \ref{fig:audio_source_distance_tube_length} indicates that the attacker can use an audio source from a larger distance if the sampling tube length is shorter or no sampling tube is present. For example, the audio source can generate a 2.5 Pa forged pressure at 7 cm far from the pressure ports without a sampling tube (Fig. \ref{fig:audio_source_distance_tube_length} (right)). Sampling tube length depends on the location of DPSs from the pressure ports. Depending upon different locations of DPSs, the sampling tube length can be very short, or even no sampling tube can be present. The attacker can target those DPSs for greater impact.


\textbf{2. LPF and the resonant frequency:} Section \ref{subsec:Electronics inside of a DPS} mentions that a DPS has an LPF. Therefore, simply filtering the resonant frequency using an LPF can prevent the resonance in DPS. However, manufacturers don't use the LPF to filter out the resonant frequency because the resonant frequency of a DPS is not constant. A resonant frequency not only depends on the transducer and diaphragm of the DPS but also depends on the sampling tube's length and diameter, the fluid's viscosity and density inside of the sampling tube (see Sections \ref{subsec:Factors not influencing the resonant frequencies} and \ref{subsec:Resonance with sampling tube in NPRs}). Therefore, it \textbf{varies} within a \textit{band} for different transducer systems depending upon different applications. Moreover, manufacturers also don't filter out the whole \textit{band} where the resonant frequency may belong. The reason is that a DPS is not only used in NPRs but also used in other \textit{dynamic pressure sensing} applications where removing a frequency band might remove important information from the input data. 

We can find a simple proof of this concept in Table \ref{table:SensorSpecAndResonantFreq}. Both of the digital DPSs in Table \ref{table:SensorSpecAndResonantFreq} have $\sim$ 2.1 kHz sampling frequency and 760-890 Hz resonant frequency. If the LPF in the DPS filtered out the resonant frequency, we would not find the resonance.

\vspace{-0.30em}
\subsection{Limitations}
\label{subsec:limitation}

In this paper, the introduced adversarial control does not offer fine-grained control compared to \cite{trippel2017walnut,tu2018injected}. The reason behind this is that the direct feedback from the compromised NPR to the attacker is absent. Because, typically, the audio sources, such as cellphones, radios, televisions, and CCTVs, which inject malicious music, do not have pressure sensors to measure the pressure after the attack and send it back to the attacker. However, the attack is strong enough to change the negative pressure in an NPR. Moreover, close access near the pressure ports in an NPR, short-attacking range, and prior knowledge of the NPR are also the limitations of our attack model.


\vspace{-0.30em}
\section{Countermeasures}
\label{sec:Countermeasures}

The following techniques should be adopted together to prevent our attack - a wolf in sheep's clothing.

\begin{figure}[h!]
\vspace{-0.60em}
\centering
\includegraphics[width=0.4\textwidth,height=0.15\textheight]{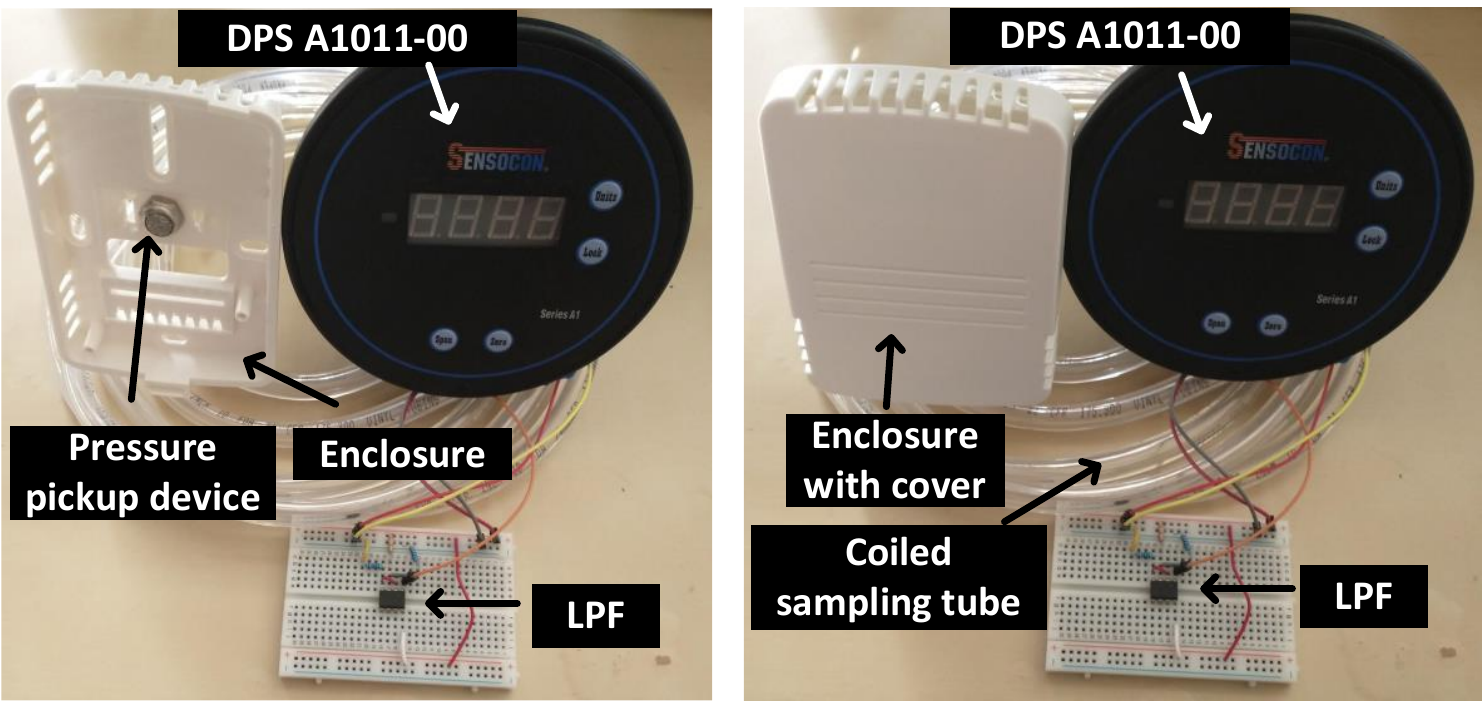}
\vspace{-1.20em}
\caption{Different countermeasures to prevent the attack.}
\label{fig:countermeasure}
\vspace{-01.10em}
\end{figure}

\textbf{Dampening of the music:} The simplest method of preventing resonance originating from the malicious music is to dampen the music. The smart way to dampen the music is to use a long sampling tube with the DPS's port. Even if the pressure port is very close to the DPS and the DPS would not require the sampling tube, we still suggest using a long sampling tube with the DPS. We find that a tube length greater than 7 m can completely dampen music having an SPL of 90 dB. The long tube can be coiled if space is limited for the mounting (see Fig. \ref{fig:countermeasure}). However, a long sampling tube reduces the sensitivity of the DPS, resulting in a measurement error. 

\textbf{Enclosure around the pressure port:} A box-like enclosure should enclose the pressure pickup device mounted in the pressure port (see Fig. \ref{fig:countermeasure}). The box-like enclosure should be filled with sound damping foam to dampen the malicious music. However, this method also reduces the sensitivity of the DPS.

\textbf{Filtering the resonant frequency:} Though the DPSs don't use their LPFs to remove the resonant frequency, the authority of the NPR facility can ask the company, that install the RPM system or BMS, to cascade an LPF just after the DPS. The LPF must have a lower cut-off frequency, such as a frequency $\sim$20\% of the resonant frequency of a DPS (see Fig. \ref{fig:resonant_frequency}). Therefore, the variability of the resonant frequency will not impact the safety of the DPS. For example, we use a first-order LPF built with an Op-Amp having a cut-off frequency of $\sim$120 Hz with the A1011-00 DPS from Sensocon (see Fig. \ref{fig:countermeasure}). A low cut-off frequency of an LPF will not hamper the normal operation of a DPS in an NPR as the pressure does not change in high frequency in an NPR. Another complex approach is to use a microphone to sense the music first and then filter out the music from the pressure reading using an LPF. Similar techniques are found here \cite{barua2022premsat, barua2022halc, barua2022sensorsecurity, chhetri2017cross}. Moreover, a guideline should be adopted by CDC or other authorities that NPRs should strictly use LPFs to protect from the resonance in DPSs.

\textbf{Increasing the reference negative pressure:} The CDC or other authorities should have a guideline to maintain a negative pressure higher than -2.5 Pa, such as at least 20 Pa. An attacker may find it difficult to turn a high negative pressure into a positive pressure through malicious music.

\textbf{Removing audio sources:} Any audio source should be removed from the close proximity to the DPS. Even CCTVs should be mounted at least 3 m away from the pressure ports in an NPR.

\vspace{-01.1em}
\section{Related Work}

To the best of our knowledge, there is no work in the literature that shows an attack on an NPR facility using malicious music by exploiting the resonant frequency of a DPS. We compare our work with the state-of-the-art works in the following four categories. 

\textbf{Attacks on pressure Sensors:} Rouf et al. \cite{rouf2010security} used unauthenticated wireless transmission to spoof a tire pressure sensor using a radio frequency (RF) channel and attacked a moving vehicle from a close distance. 
Tu et al. \cite{tu2021transduction} showed a deliberate EMI attack on an inflation pump's pressure sensor while inflating a car tire and studied the attack impacts on the system's actuation. Yan et al. \cite{6985040} did a formal analysis of semantic attacks on pressure sensors \textit{without} mentioning how the pressure sensors can be attacked.

\textbf{Attacks with acoustic signals:} Wang et al. \cite{wang2017sonic} used an ultrasonic gun to create resonance at membranes of different inertial sensors, such as MEMS accelerometers and gyroscopes and spoofed the inertial sensors to create havoc in the connected systems. Son et al. \cite{son2015rocking} used a high-power acoustic signal in audible range to compromise the gyroscope of a drone creating a resonance and made it uncontrollable. Trippel et al. \cite{trippel2017walnut}, and Tu et al. \cite{tu2018injected} showed an adversarial control over MEMS accelerometers and gyroscopes using audible acoustic signals at their resonant frequencies. Yan et al. \cite{yan2016can} showed an attack on ultrasonic sensors of a vehicle using acoustic waves to impair vehicle safety. Zhang et al. \cite{zhang2017dolphinattack} injected acoustic commands into a microphone using ultrasonic carriers. Bolton et al. \cite{bolton2018blue} showed an acoustic attack on hard disk drives.

\textbf{Resonant frequencies in pressure sensors:} The resonant frequency of a pressure sensor influences its dynamic characteristics \cite{hjelmgren2002dynamic} and is a critical parameter in designing a pressure sensor. Designers use this frequency to design resonant pressure sensors for dynamic applications, such as \cite{han2020novel}, \cite{greenwood1988miniature}, and \cite{shen2019cylinder}. We are not aware of any acoustic attack on pressure sensors exploiting resonant frequencies. However, designers design pressure sensors to acquire acoustic pressure in different applications, such as for cardiac pressure \cite{wang2021acoustic} and sound pressure \cite{nagiub1999characterization}. 

\textbf{Attacks on other sensors:} Barua et al. \cite{barua2020hall, barua2021hall, barua2020special} showed a non-invasive magnetic spoofing attack on Hall sensors of solar inverters, causing a shut down in a micro-grid. Kune et al. \cite{kune2013ghost} attacked analog sensors using EMIs to cause defibrillation shocks on implantable cardiac devices. Davidson et al. \cite{davidson2016controlling} showed how spoofing optical sensors of an unmanned aerial vehicle (UAV) can compromise complete control of its lateral movement.

While the above works address the physical-level signal injection attacks on different sensors, our work differs from them in the following ways. \textbf{First}, our attack is the first of its kind that exploits resonant frequencies of DPSs to attack the RPM and HVAC systems in an NPR facility. \textbf{Second}, we intelligently use malicious music to attack NPRs for stealthiness (i.e., a wolf in sheep's clothing). \textbf{Last}, more importantly, our attack has the potential to trigger catastrophic consequences by leaking deadly microbes from an NPR, causing losses in terms of human lives and monetary resources.

\section {Conclusion}
\label{sec:conclusion}

We present a non-invasive attack using malicious music on DPSs located in an NPR. We show that the NPRs have RPM and HVAC systems, which use DPSs to maintain a negative pressure inside an NPR with respect to the outside reference space. We find the resonant frequency of DPSs used in NPRs by proper experiments and show that the resonant frequencies are in the audible range. We also show that the resonant frequencies of DPSs vary within a band depending on other parameters, such as the length and diameter of the sampling tube. Therefore, we insert segments of the resonant frequency band in specific interval inside of music and end the inserted segments with their peak to maintain an average forged pressure in the DPS's transducer system. As a result, the attacker can use the malicious music to fool the DPSs used in the RPM and HVAC systems of an NPR and can turn the NPR's negative pressure into a positive pressure. This may cause an alarm, resulting in chaos in the facility and has a potential to leak deadly microbes from the facility. Our attack is strong, non-invasive, and stealthy, similar to a wolf in a sheep's clothing. The consequences of leaking deadly microbes from an NPR will be catastrophic in terms of losses in human lives and monetary resources. Therefore, our attack is impactful, and the countermeasures should be adopted to prevent any future attack like ours in an NPR.

\begin{acks}

The authors would like to thank the anonymous reviewers for their valuable comments that greatly helped to improve this paper. This work was partially supported by the National Science Foundation (NSF) under award ECCS-2028269 and the University of California, Office of the President award LFR-18-548175. Any opinions, findings, conclusions, or recommendations expressed in this paper are those of the authors and do not necessarily reflect the views of the funding agencies.

\end{acks}


\section{Appendix}
\label{sec:appendixA}

\subsection{Types of differential pressure sensors}
\label{appendix:Types of DPSs}

\begin{figure}[ht!]
\vspace{-0.50em}
\centering\includegraphics[width=0.28\textwidth,height=0.14\textheight]{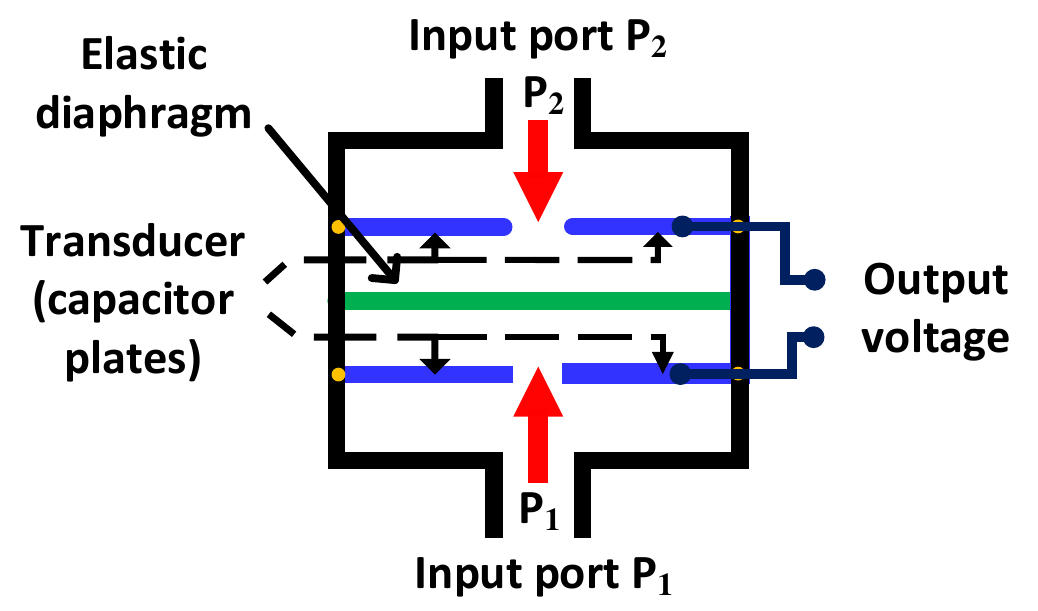}
\vspace{-1.0em}
\caption{A capacitive transducer based DPS.}
\label{fig:capacitive_DPS}
\vspace{-0.280em}
\end{figure}

\textbf{Capacitive DPS}: It uses a diaphragm placed in between the rigid plates of a capacitor that is shown in Fig. \ref{fig:capacitive_DPS}. The diaphragm works as a partition between two ports - port 1 and port 2, of the DPS.  If port 1 is in pressure level $P_1$ and port 2 is in pressure level $P_2$, the diaphragm changes its shape in proportion to the amount of differential pressure $P_1$ - $P_2$ applied to it. The change of shape of the diaphragm changes  the capacitance of the capacitor. The change of capacitance generates a proportional voltage at the sensor output.

\begin{figure}[ht!]
\vspace{-0.50em}
\centering\includegraphics[width=0.35\textwidth,height=0.12\textheight]{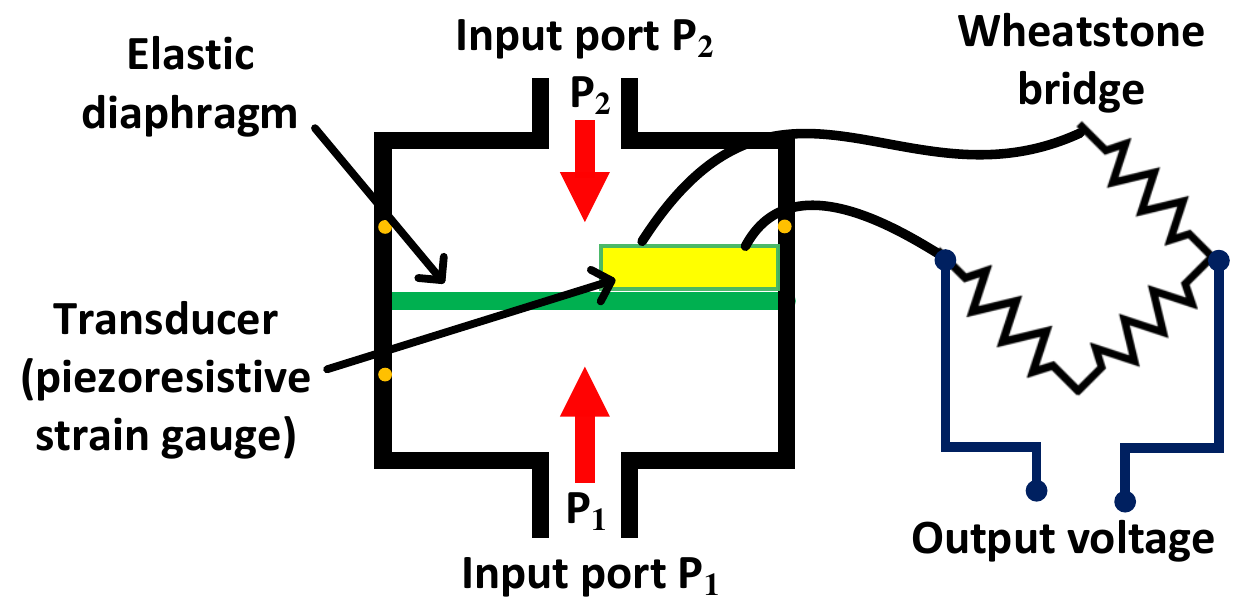}
\vspace{-1.15em}
\caption{A piezoresistive transducer based DPS.}
\label{fig:piezoresistive_DPS}
\vspace{-01.280em}
\end{figure}

\textbf{Piezoresistive DPS}: It uses a piezoresistive strain gauge as a transducer that is connected with a diaphragm (see Fig. \ref{fig:piezoresistive_DPS}). As the diaphragm is placed in between two ports of the DPS, the  diaphragm's shape changes in proportion to the differential pressure  $P_1$ - $P_2$ applied on the diaphragm, causing a change in shape of the piezoresistive element connected to the diaphragm. This changes the resistance of the piezoresistive element, which is typically used as an arm of a Wheatstone bridge.  Therefore, the change in resistance results in  a proportional voltage change at the sensor output.

\begin{figure}[ht!]
\vspace{-0.70em}
\centering
\includegraphics[width=0.3\textwidth,height=0.11\textheight]{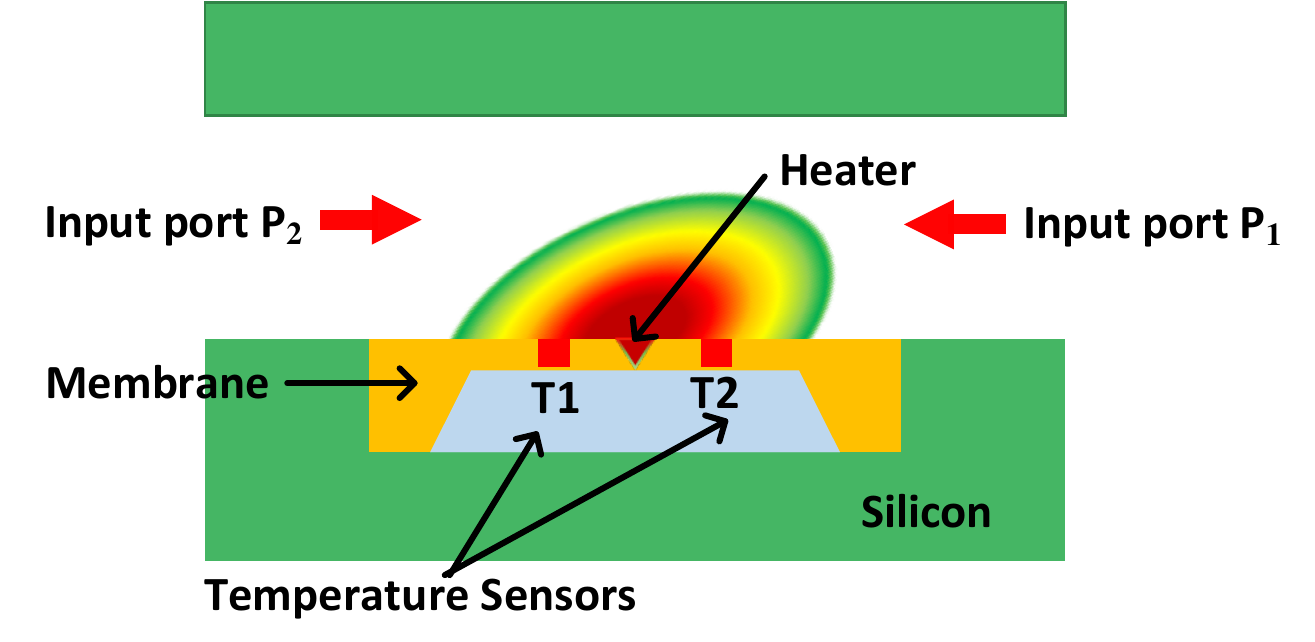}
\vspace{-0.81em}
\caption{A thermal mass-flow based DPS.}
\label{fig:thermal_DPS}
\vspace{-01.2800em}
\end{figure}

\textbf{Thermal mass-flow DPS}: It uses temperature sensors as  transducers and can measure differential pressure utilizing the thermal gas-flow principle \cite{datasheetsdpTechnology}. As shown in Fig. \ref{fig:thermal_DPS}, it has two temperature  sensors $T_1$ and $T_2$, and  a small  heating  element is placed in the middle of the temperature sensors. The structure is etched into a passivation glass layer, which forms a thin membrane.  A differential pressure across the sensor ports $P_1$  and $P_2$ induces a tiny gas flow, which results a temperature difference $T_1$ - $T_2$ between the two temperature sensors. The temperature difference results in a proportional voltage change at the sensor output.

\subsection{Signal conditioning circuit}
\label{appendix:Signal_conditioning_circuit}

\begin{figure}[h!]
\vspace{-00.0em}
\centering
\includegraphics[width=0.4\textwidth,height=0.14\textheight]{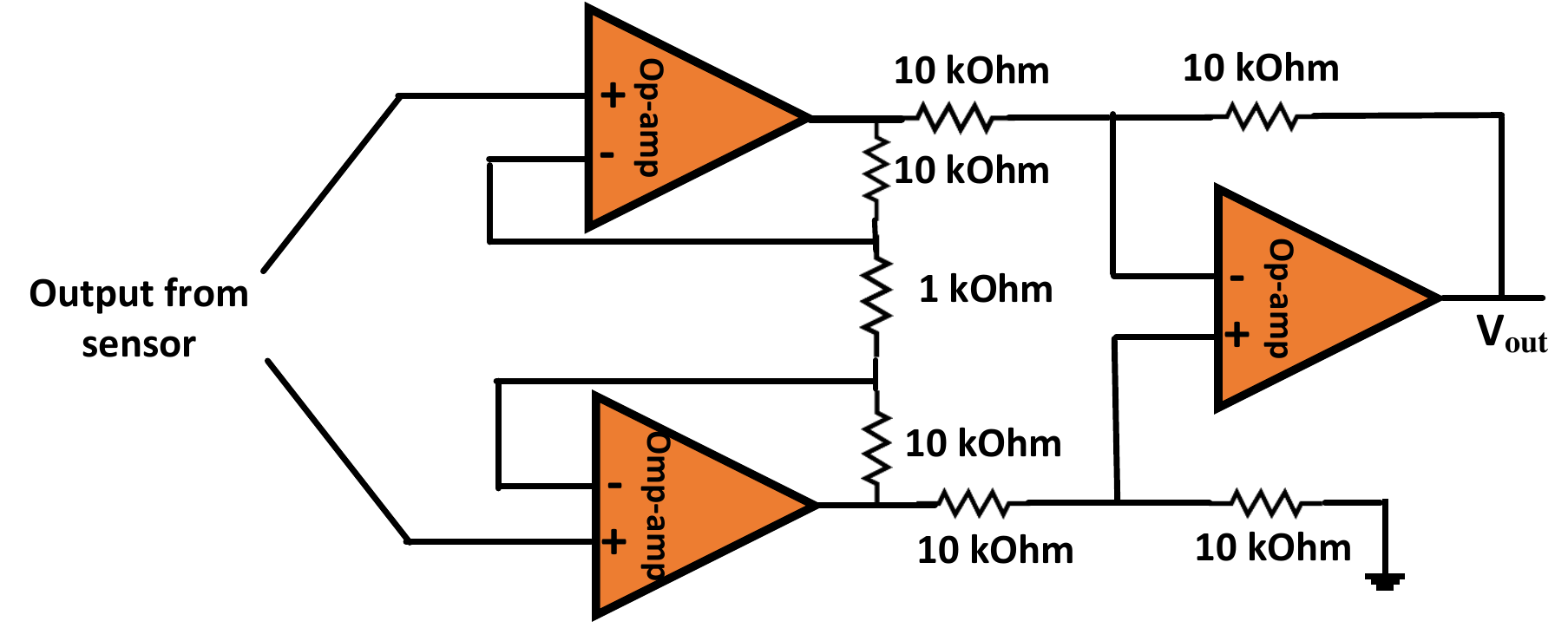}
\vspace{-0.5em}
\caption{Instrumentation amplifier.}
\label{fig:signal_conditioning_circuit}
\vspace{03.0em}
\end{figure}

Fig. \ref{fig:signal_conditioning_circuit} shows an instrumentation amplifier to collect data from a DPS  with the part\# NSCSSNN015PDUNV.

\Urlmuskip=0mu plus 1mu\relax
\bibliographystyle{ACM-Reference-Format}
\balance
\bibliography{ccs-sample}

\end{document}